\documentstyle[aps,prd,epsf,floats,amsmath,amssymb]{revtex}

\newcommand{\bk}{{\bf k}}
\newcommand{\bq}{{\bf q}}
\newcommand{\bzero}{{\bf 0}}

\title{Sigma and hydrodynamic modes along the critical line}
\author{H.~Fujii$^a$\footnote{Email: hfujii@phys.c.u-tokyo.ac.jp}
    and M.~Ohtani$^b$\footnote{Email: ohtani@rarfaxp.riken.jp} }
\address{%
$^a$Institute of Physics, University of Tokyo, Tokyo 153-8902, Japan\\
$^b$Radiation Laboratory, RIKEN,
Wako, Saitama 351-0198, Japan}
\date{\today}

\begin{document}
\draft
\maketitle

\begin{abstract}
Assuming a tricritical point of the two--flavor QCD in the space of
temperature, baryon number chemical potential and quark mass, we study
the change of the associated soft mode along the critical line within the
Ginzburg--Landau approach and the Nambu--Jona-Lasinio model. 
The ordering density along the chiral critical line is the scalar
density whereas a linear combination of the scalar, baryon number and
energy densities becomes the proper ordering density along the
critical line with finite quark masses. 
It is shown that the critical eigenmode shifts from the sigma--like
fluctuation of the scalar density to a hydrodynamic mode at the
tricritical point, where we have two ordering densities, the scalar
density and a linear combination of the baryon number and energy densities.
We argue that 
appearance of the critical eigenmode with hydrodynamic character
is a logical consequence of divergent susceptibilities of the conserved 
densities.

\end{abstract}


\section{Introduction}

At high temperature and/or baryon density,
the system governed by QCD will show a transition from
an ordinary hadronic phase to a chirally symmetric,
deconfined plasma phase\cite{SY82,PW84}.
The main objective of the heavy--ion programs at RHIC and at
future LHC 
is to create this long-sought plasma state and to
study collective 
properties of this many--body assembly\cite{QGP3}.
These two phases would have to be separated 
by a boundary with singularity
if chiral symmetry and/or confinement of QCD were exact symmetry.
In reality, dynamical quarks with finite masses 
$m \ne 0$
make both symmetries only approximate, and 
their order parameters, the quark condensate and the Polyakov loop,
have non--vanishing values everywhere in the phase diagram.
Thus the plasma state may be smoothly connected with the
ordinary hadronic state,
even though they would possess qualitatively
different properties from each other.

Recently a strong possibility of a critical point
in the real QCD phase diagram was suggested\cite{RS98,RS99,MAS04},
based on model 
calculations\cite{AY89,BCDGP89,BCPG94,GGP94,BR99,HJSSV98,SMMR01,HI03}
as well as lattice QCD results\cite{FK02,FP03,KAE03}.
It is the endpoint of the first--order line, 
inferred from the crossover behavior
along the temperature ($T$) axis and the first order transition along the
axis of the baryon--number chemical potential ($\mu_B$),
and is a genuine singular point with 
the same criticality as the Z$_2$ Ising model.
Its location,
which is sensitive to the strange quark mass $m_s$,
is expected to be 
within the reach of current experimental facilities.
Observable implications of this Z$_2$ critical point (Z$_2$CP)%
\footnote{In the QCD thermodynamics this point is
referred to as critical end point (CEP).
In this paper we use Z$_2$CP instead of CEP, to indicate the
symmetry of the point.}
in heavy ion experiments have been discussed in the 
literature\cite{RS98,RS99,MAS04,BR00,BPSS01,BG01,PSD03,AFK03,HS03}
such as large fluctuations of the low--momentum
particle distributions, and the limitations on them
due to the finite space--time geometry of collision events.
This  Z$_2$CP
will become a critical cornerstone in the QCD phase diagram
once its location is confirmed in experiments.

Based on the approximate chiral symmetry, 
the scalar density is usually taken as the order parameter
of the Ginzburg--Landau (GL) effective potential
to describe the critical behavior at the Z$_2$CP.
In this description all the singularities
associated with the Z$_2$CP seemingly
originate from softening of the scalar density fluctuations
as the effective potential becomes flat there.
Especially, it might be concluded that the sigma meson becomes
massless as an immediate consequence of this critical point.

As a basic fact, however, we should strictly distinguish between
the chiral critical point with $m=0$
and the Z$_2$CP with $m\ne 0$ ---
even within the chiral effective models.
When the chiral symmetry is exact, the $T$--$\mu_B$ plane is 
divided into two domains
of the symmetric and broken phases with
a boundary {\em line}.
But the symmetry argument is unable to fix the order of the singularity
of this line,
especially the possible existence of
the tri--critical point (TCP) on this line.
Since the Z$_2$CP at finite quark mass is the remnant of this TCP,
the relation of the Z$_2$CP to the chiral symmetry is very obscure.
In fact, Z$_2$ is not the symmetry
of the underlying interactions, but of the
thermodynamic potential at this particular isolated {\em point}
in the $T$--$\mu_B$ phase diagram\cite{KLS01}.
From this point of view the Z$_2$CP is different from
the chiral critical point.

The flat effective potential indicates
the large fluctuation of the scalar density.
At the chiral critical point,
this is related to softening of the sigma meson
mode, so as to form a symmetric excitation
spectrum together with the pions.
On the other hand, there is no such symmetry constraint at the Z$_2$CP.
Actually the scalar fluctuation 
{\em linearly} mixes with fluctuations of baryon number density
and energy density,
and therefore not only the sigma mode
but also the hydrodynamic mode are to be taken into account 
there to study the associated soft mode.
Consequently the fate of the sigma meson mode at the
Z$_2$CP is non--trivial.

Recent calculations of the dynamic mode in the scalar
channel using the chiral models\cite{SMMR01,HF03} 
indeed showed that the sigma meson is massive at the Z$_2$CP.
Furthermore, 
another scalar mode with space--like momentum dispersion is 
identified as the soft mode associated with the Z$_2$CP
in the Nambu--Jona-Lasinio (NJL) model\cite{HF03}.
In this paper we shall confirm
the result of Ref.~\cite{HF03}
on the more general ground 
using the time--dependent Ginzburg--Landau (TDGL) approach,
and extend the study to discuss the changeover of the
soft modes along the critical line 
in the $T$--$\mu_B$--$m$ space
within the TDGL approach as well as the NJL model.

Our investigation is based on two fundamental observations about
the Z$_2$CP. The first point is that the proper
ordering density at the Z$_2$CP is a linear combination
of the scalar, baryon number and internal energy 
densities\cite{KLS01,HF03,FO04}, as mentioned above.
Because of this mixing
all the susceptibilities of these densities
diverge with the same critical exponent at the Z$_2$CP. 
In contrast, in the chiral critical transition,
the susceptibility of the scalar density diverges with 
exponent $\gamma$ of the O(4) model in the two--flavor case,
while the other susceptibilities
of the baryon number and the energy have the smaller exponent $\alpha$.

The second is a consequence on the dynamics following 
from the conservation of the baryon number and the
energy.
The fluctuations of these conserved 
densities\footnote{%
Momentum density is neglected here for simplicity.}
are intrinsically soft 
and
constitute the hydrodynamic modes,
whose excitation energies vanish as the wavevector $\bq$ goes to zero.
Susceptibilities of these conserved densities in turn
have the spectral contributions solely from these
hydrodynamic modes when expressed as
a sum of mode spectra\cite{DF75,TK91,CMT03,FO04}.
Hence the divergence of the susceptibility of a conserved density
must be accompanied by critical slowing of 
a hydrodynamic mode.
The spectral contribution from this 
hydrodynamic mode may well be involved in the
scalar susceptibility through the mixing at the Z$_2$CP.

At an O(4) critical point (O(4)CP) the importance of the
hydrodynamic mode 
depends on which phase we start from.
The hydrodynamic mode plays no critical 
role in the symmetric phase whereas the scalar condensate
makes the mixing possible in the broken phase.
The situation becomes  more subtle at the
TCP, where the O(4) critical line
shifts to the first--order line. 
Only the scalar susceptibility
diverges due to the softening of the sigma meson
at the TCP if it is approached from the symmetric phase. 
Otherwise, the hydrodynamic soft mode
causes the divergence in
the susceptibilities of the baryon number and energy
as well as the scalar one.

This paper is organized as follows. In the next section we briefly
review generic properties of the
phase diagram of QCD with two flavors near the TCP
using the GL effective potential.
It is stressed that at the TCP 
there are two relevant order
parameters, 
the scalar condensate and a conserved density
which is a linear combination of the baryon number and
entropy densities.
Then we include the dynamics using the TDGL model.
Writing the susceptibilities as a spectral sum,
we discuss the relative weight of the spectral contributions from the
sigma and hydrodynamic modes.
It is pointed out that the hydrodynamic contribution generates the
discontinuity of the baryon number and 
entropy susceptibilities at the O(4)CP,
and that this hydrodynamic mode gives the divergence at the
TCP approached from the broken phase and also at the Z$_2$CP.
In \S III we perform the same analysis using
the NJL model as an illustration. The GL effective potential with
two ordering densities are numerically constructed there.
The flat directions at the critical points are shown and discussed
in relation to the divergences of the susceptibilities.
The spectral origins of these divergences are studied
with the relative weight of the mode spectra, and 
in detail based on the poles and residues of the
scalar response function.
Sections IV and V are devoted to discussions and
summary. In Appendix A we prove the
relation between the susceptibility and the response function,
and in Appendix B we present the explicit formulas of the
response functions in the NJL model. The results with the chiral quark model
is briefly reported in Appendix~C.

\section{Generic analysis} 
\label{sec:2}

\subsection{Structure of the phase diagram and order parameters}

Let us briefly review the phase structure near the 
TCP%
\cite{LS84,RS99,HI03}.
It is known that the critical properties near the TCP are
described, up to logarithmic corrections,
with the Ginzburg--Landau effective potential
\begin{eqnarray}
\Omega &=&  \Omega_0(T,\mu_B) +
\int d^3 x \left (
a(T,\mu_B) \sigma^2 + b(T,\mu_B)\sigma^4 + 
c(T,\mu_B) \sigma^6 -h\sigma \right)
\nonumber \\
& \equiv& 
\Omega_0(T,\mu_B) +\int d^3 xf(T,\mu_B,h;\sigma),
\label{eq:GLeffpot}
\end{eqnarray}
where $f$ ($\Omega_0$)  denotes the (non-)singular part of
the effective potential,
and $c>0$. The pseudo--scalar density is set to zero and
neglected here in the mean field approximation.
The critical exponents can be easily
found from (\ref{eq:GLeffpot}) at the mean field level. 
Along the line of the first--order transition within the
symmetry plane ($h=0$),  we have
\begin{equation}
f=a \sigma^2 + b\sigma^4 + 
c \sigma^6 \equiv c\sigma^2(\sigma^2-\sigma_1^2)^2,
\end{equation}
where three minima with $\sigma=0,\pm \sigma_1(T,\mu_B)$
coexist (dashed line in Fig.~1).
The baryon number and entropy densities are functions of $\sigma^2$ due to
symmetry, and discontinuous across the boundary
between the symmetric phase ($\sigma=0$)
and the broken phase ($\sigma=\pm \sigma_1$). 
At the TCP, 
where $a=b=0$, these {\em three} phases coalesce and the first order
line ($b=-2\sqrt{ac}$) smoothly joins with the O(4) critical 
line ($a=0, b>0$).

Once a small explicit breaking field $-h\sigma$ 
is exerted,
the O(4) critical line disappears and the TCP is lifted
to the Z$_2$CP.
The line of Z$_2$CP as a function of $h$
is determined by the
condition, $ f'=f''=f'''=0$ ($'\equiv \partial/\partial \sigma$),
which is solved for a negative $b$ with
$a=3b^2/5c$, $b=-\sigma^2/5$ and $\sigma={\rm sign}(h)(|h|/16)^{1/5}$.
Two lines of Z$_2$CP with $h \gtrless 0$
form the edge of 
the wing-like surface of the first order transition 
in the $a$--$b$--$h$ space, and 
these lines connect smoothly to
the O(4) critical line at the TCP.
This wing structure is mapped into the physical phase space 
of $T$, $\mu_B$ and $h\sim m$ (see Fig.~1).

The slope of the first--order boundary can be related to the 
discontinuities of the densities across the boundary via
the Clapeyron--Clausius relation\cite{BCPG94,HJSSV98},
\begin{equation}
\frac{dT}{d\mu_B}=-\frac{\Delta \rho_B}{\Delta s},\quad
\frac{dT}{dh}=-\frac{\Delta \sigma}{\Delta s},\quad
\frac{dh}{d\mu_B}=-\frac{\Delta \rho_B}{\Delta \sigma}
\label{eq:CCrel}
\end{equation}
with  baryon number density $\rho_B$  and entropy density $s$.
The chiral broken phases with
$\sigma=\pm \sqrt{-a/2b}$ coexist within the symmetry plane $h=0$,
and accordingly there is no gap in 
$\rho_B$  and $s$ across this symmetry plane.
Only the scalar density $\sigma$ bifurcates as the ordering density
at the O(4)CP approached from 
the symmetric phase.
Its correlations with the ``energy--like'' densities vanish 
$\langle \sigma s \rangle=\langle \sigma \rho_B \rangle=0$
because of the symmetry in $\sigma \leftrightarrow -\sigma$
in the symmetric phase.

\begin{figure}[tb]
\begin{center}
\centerline{\epsfxsize=0.55\textwidth
\epsffile{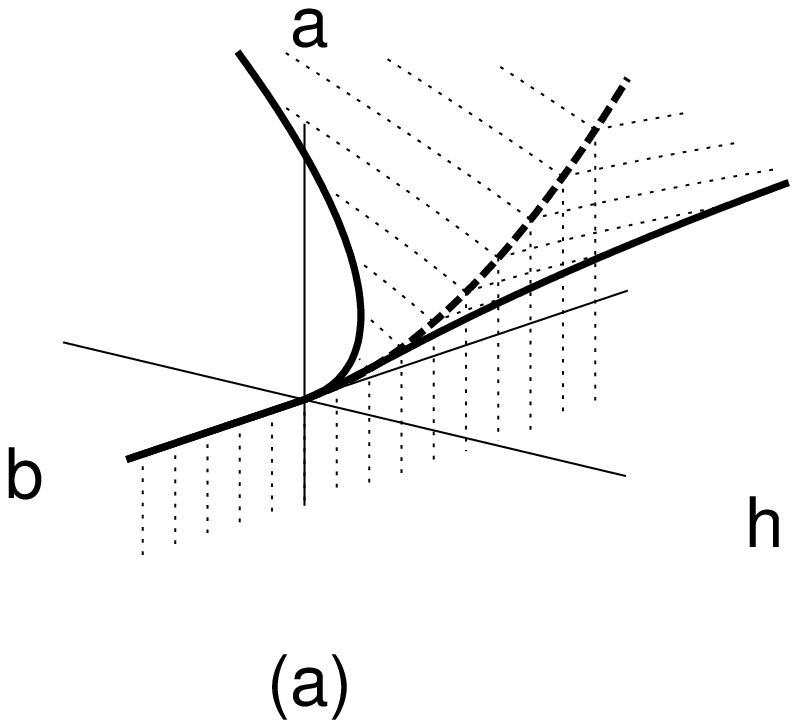}
\hspace{-15mm}
\epsfxsize=0.45\textwidth
\epsffile{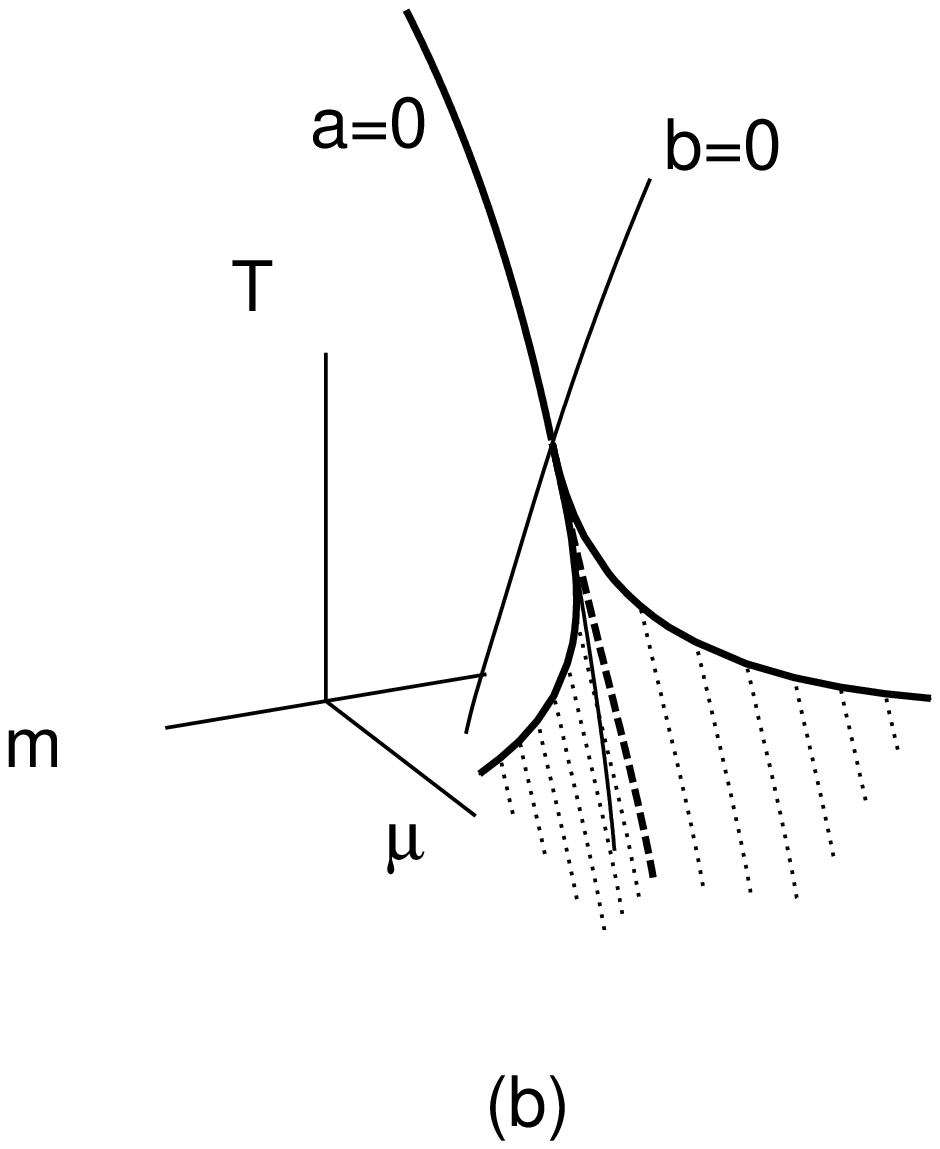}
\hfill}
\caption{(a) Schematic phase diagram around the 
TCP in the $a$--$b$--$h$ space.
The three critical lines are shown in bold lines which meet
at the origin (TCP). The curve of three--phase coexistence
is drawn
in a bold dashed line which ends at TCP. 
The two-phase coexistence surface
are hatched by thin dotted lines.
(b) The counterpart in the physical 
$T$--$\mu$--$m$ space (NJL model).}
\end{center}
\end{figure}

From the relation (\ref{eq:CCrel}) we know that 
all of $\sigma$, $\rho_B$ and $s$ generally have
discontinuities across the
wing because there is no reason for any of these
slopes to vanish once $h\ne0$.
Let us discuss the energy--like and ordering densities around 
the Z$_2$CP\cite{KLS01,WN96}.
First we introduce the ``temperature--like''
field as a vector tangential to the
coexistence boundary.
Then the energy-like density is defined
as the thermodynamic variable conjugate
to this temperature--like field.
This density has no
 discontinuity in the vicinity of the critical point.
Since the boundary is two--dimensional,
there are two independent temperature--like fields 
and correspondingly two energy--like densities.
Next the ordering density is defined as the density whose correlations
with the energy--like densities vanish
at the critical point approached from the ``symmetric'' phase
along the temperature--like direction.
The conjugate field of this ordering density is no longer
tangential to the coexistence boundary.
There is a single ordering density at the Z$_2$CP,
which is in general a linear combination of
$\sigma$, $\rho_B$ and $s$.
Since all the susceptibilities of these densities
include the same singular fluctuation,
they diverge with the same critical exponent at the Z$_2$CP.

The coexistence wing is squeezed to be one--dimensional
at the TCP, where 
two lines of the Z$_2$CP and the line of the first order
transition with $h=0$ merge and smoothly connect to the single
O(4) critical line.
Thus at the TCP we have only one energy--like density, which will be
a linear combination of $\rho_B$ and $s$.
Accordingly there are {\em two} ordering densities from dimensionality.
The obvious one is the scalar density $\sigma$
related to the
chiral symmetry and the other $\varphi$
is 
another linear combination of $\rho_B$ and $s$  representing the 
Z$_2$ symmetry of the potential at this particular point. 
It is sometimes useful to construct the effective potential 
with two ordering densities, 
$\sigma$ and $\varphi$, which become soft at the TCP.

The same observation can be made by looking at the
susceptibilities directly.
There are three fields $h$, $a$ and $b$ 
in the effective potential (\ref{eq:GLeffpot}).
The singular parts of the corresponding susceptibilities form
a 3-by-3 matrix
($i,j= h, a, b$),
\begin{equation}
\chi_{ij}=
\chi_{h}
\left (
\begin{array}{ccc}
1         & 2\sigma     & 4 \sigma^3 \\
2\sigma   & 4 \sigma^2  & 8 \sigma^4 \\
4\sigma^3 & 8 \sigma^4  & 16 \sigma^6
\end{array} \right ) 
\label{eq:chimatrix}
\end{equation}
with $\chi_{h}$ the scalar susceptibility,
\begin{eqnarray}
\chi_{h} &=& -{1 \over V}
\frac{\partial^2 \Omega }{ \partial^2 h}
=\frac{1}{2 a + 12  b \sigma^2 + 30 c \sigma^4},
\label{eq:chih}
\end{eqnarray}
where $\sigma$ takes the value at the extremum of the potential.
When the O(4)CP is approached from the symmetric phase,
$\chi_{aa}=\chi_{bb}=0$.
In the broken phase the situation is different.
The singular part of (1) gives a finite contribution 
$\chi_{aa}=1/(3b)$
as the O(4)CP is approached with $\sigma^2  = -a/2b\to 0$,
although the divergent susceptibility is still the scalar one alone.
The singular contribution to $\chi_{aa}$
eventually blows up at
the TCP approached along the O(4) critical line.
In fact, 
the TCP may be understood as a usual critical point with 
the ordering density $\varphi$ conjugate to $a$,
sitting on the
chiral phase boundary.
When the TCP is approached from the broken phase
 with $b=h=0$, the scalar ordering
density vanishes slowly $\sigma^4  = -a/3c$, and
$\chi_{h}$ and $\chi_{aa}$ diverge like
$1/|a|$ and $1/\sqrt{|a|}$, respectively, while $\chi_b$ is 
still non-singular.  
Note that the divergence of $\chi_{aa}$ 
at the TCP indicates
the infinities in the baryon number and energy%
\footnote{The fluctuation of the energy density is a linear combination of those
of the baryon number and entropy densities. In this paper we sometimes
use the energy susceptibility and the entropy susceptibility interchangeably.}
susceptibilities, or
equivalently the isothermal 
compressibility and specific heat, respectively.

All these susceptibilities in (\ref{eq:chimatrix})
diverge at the Z$_2$CP, where $\sigma \ne 0$.
With the finite condensate $\sigma$
we can diagonalize this matrix of the susceptibilities,
leaving only one singular susceptibility.
The resulting eigenvalues are 
(0,0,$(1+4 \sigma^2+16 \sigma ^6)\chi_h$),
with eigenvectors
$^t(-2 \sigma,1,0)$,
$^t(-4 \sigma^3,0,1)$,
$^t(1,2\sigma,4\sigma^3)$, respectively.
For small $\sigma$ or $h$, we see that
the ordering density is approximately
a linear combination of the densities $\sigma$ and $\varphi$.

The fluctuations of these
two ordering densities $\sigma$ and $\varphi$
become large near the TCP as explained above, 
and should
be included as the soft degrees of freedom,
especially when we discuss the dynamic aspects.
We generalize the free energy so as to
have two ordering densities,
\begin{equation}
\Omega = \int d^3 x \left (
a_0 \sigma^2 + b_0 \sigma^4 + c\sigma^6 
+\gamma \sigma^2 \varphi + \frac{1}{2}\varphi^2
-h \sigma -j \varphi  \right )
+\Omega_0.
\label{eq:GL2var}
\end{equation}
Coupling between $\sigma$ and
$\varphi$ must respect the underlying chiral symmetry
and the simplest coupling is $\sigma^2 \varphi$.
A flat direction of this potential appears
at a critical point in the $\sigma$--$\varphi$ plane.
In the case of the O(4)CP/TCP it is in the $\sigma$ direction
reflecting the symmetry while the direction will become
a linear combination of the two densities at the Z$_2$CP.
Eliminating the density $\varphi$ by
$\partial \Omega/\partial \varphi=
\gamma \sigma^2 + \varphi -j=0$,
we recover the original form (\ref{eq:GLeffpot})
of the free energy 
with $a=a_0+\gamma j$ and $b=b_0-\frac{1}{2}\gamma^2$,
up to an analytic term.

\subsection{Dynamics}

We may introduce the dynamics to the system described by the
free energy (\ref{eq:GL2var}) phenomenologically\cite{HH77}.
We have seen in the previous subsection that
there are two ordering densities conjugate to the fields
$h$ and $a$ at the TCP, and that a {\em linear}
combination of these two densities
will become the relevant ordering density at the Z$_2$CP
for small $h$ or $\sigma$.
We should include at least these two densities in order to
describe the soft dynamics.
Furthermore, it is known that (non--linear) 
mode--mode coupling between
the fluctuations of the ordering densities  and
other (non-critical) hydrodynamic modes are
important in general to
describe the dynamics in the critical region\cite{HH77},
which is beyond the scope of this work\cite{SS04,KO04}.
We will see, however, that
the coupled system of the two ordering densities
in the mean field approximation yields already a non-trivial 
result\cite{FO04}.

\subsubsection{Mixing between  scalar and conserved densities}

Deviation of the densities from the
absolute equilibrium gives rise to time evolution of the system.
Here we assume simple phenomenological 
equations of motion for densities $\sigma$ and $\varphi$ as  
\begin{eqnarray}
L_\sigma(i\partial_t)\sigma
&=&- \frac{\delta \Omega}{\delta \sigma},\quad
L_\varphi(i\partial_t) \varphi
=- \frac{\delta \Omega}{\delta \varphi},
\end{eqnarray}
where $L_\sigma(i\partial_t)$ and $L_\varphi(i\partial_t)$ 
are the differential operators.
Appropriate  forms of $L_\sigma$ and $L_\varphi$ are unknown 
in this description. But 
as a strong constraint we know that 
the operator $L_\varphi$ must be consistent with the conservation of the
density $\varphi$ and describe the hydrodynamic motion.
As a typical hydrodynamic evolution,
we consider here the diffusion motion
$L_\varphi(i\partial_t)=-\partial_t/\lambda \bq^2$
with wavevector $\bq$. 
Note that the diffusion is time--irreversible.
We assume propagating motion 
$L_\sigma(i\partial_t)=\partial_t^2/\Gamma$
for the scalar density, identifying
this mode as the sigma meson which degenerates 
with the propagating pions at the O(4)CP/TCP.
Other possible forms are considered below in this section.
The transport coefficients $\Gamma,\lambda>0$
are treated as constants here.

For small deviations
$\sigma \to \sigma+\tilde \sigma$ and 
$\varphi \to \varphi+\tilde \varphi$
from the equilibrium values,
we linearize these equations of motion
with respect to $\tilde \sigma$ and $\tilde \varphi$
to obtain
\begin{eqnarray}
\left (
\begin{array}{cc}
L_\sigma(i\partial_t)+\Omega_{\sigma\sigma}    &
 \Omega_{\sigma\varphi} \\
\Omega_{\sigma\varphi}     &
L_\varphi(i\partial_t)+\Omega_{\varphi\varphi}
\end{array} \right )\left(
\begin{array}{c}
\tilde \sigma \\ \tilde \varphi 
\end{array}\right)
=0,
\end{eqnarray}
where $\Omega_{\sigma\sigma}
=\delta^2 \Omega/\delta \sigma \delta \sigma |_{\rm eq}$, etc.
The soft eigenmodes of the system
are determined by the condition
\begin{eqnarray}
\left |
\begin{array}{cc}
-\omega^2+\Gamma(\chi_h^{-1}+4\gamma^2\sigma^2 +\kappa \bq^2) &
2\gamma \sigma\sqrt{\Gamma\lambda \bq^2} \\
2\gamma \sigma\sqrt{\Gamma\lambda \bq^2}  &
-i\omega +\lambda \bq^2
\end{array} \right | =0 ,
\label{eq:normal}
\end{eqnarray}
where $\chi_h$ is the scalar susceptibility given in (\ref{eq:chih}),
and we introduced a term $\kappa(\nabla \sigma)^2/2$ in $\Omega$.
The eigenmodes for  small $\bq$ are found as $\omega =\pm\omega_{\rm o},
\omega_{\rm d}$ with 
\begin{eqnarray}
-{\omega_{\rm o}^2 \over \Gamma}
&=& 
-(\chi_h^{-1}+4 \gamma^2 \sigma^2 )
-\left ( \kappa + \frac{\lambda}{\Gamma}\frac{4\gamma^2 \sigma^2}
{\chi_h^{-1}+4\gamma^2 \sigma^2}
 \right ) \bq^2 ,
\nonumber \\
{-i \omega_{\rm d}\over \lambda \bq^2} &=& -  \frac{\chi_h^{-1}}
           {\chi_h^{-1}+4 \gamma^2 \sigma^2}
\equiv - \chi_j^{-1},
\end{eqnarray}
where $\chi_j$ is the susceptibility of the density $\varphi$.

The eigenmode $\omega_{\rm o}$ is oscillating
while $\omega_{\rm d}$ has the diffusion-like hydrodynamic character.
The $\omega_{\rm o}$ vanishes at the O(4)CP, being the critical
eigenmode. Although the hydrodynamic mode $\omega_{\rm d}$
is an intrinsic soft mode of the system, it does not show
the critical slowing there.
When the TCP is approached from the symmetric phase,
the situation is the same.
On the other hand, at the TCP approached from the broken phase,
both frequencies slow down,
reflecting the divergence of the susceptibilities, 
which seems reflecting the existence of two 
independent ordering densities there. 

At the Z$_2$CP the susceptibilities
$\chi_h$ and $\chi_j$ diverge with the same exponent
due to the linear mixing.
The propagating $\omega_{\rm o}$ 
is a fast mode there due to the non-zero condensate
$\sigma = (h/16)^{1/5}$,
whereas the hydrodynamic slow mode
$\omega_{\rm d}/\lambda \bq^2=-i \chi_j^{-1}$ becomes 
the critical mode ($\chi_j^{-1} \to 0$)
associated with the Z$_2$CP.
This result is similar to the 
level-crossing phenomenon where
the mode coupling makes
the lower energy mode lowered further.
The explicit $\bq^2$ factor of 
$\omega_{\rm d}$ stemming from the
hydrodynamic character results in the larger
dynamic critical exponent $z=4$ in the mean field level,
which makes $\omega_{\rm d}$ apparently slower 
than the $\omega_{\rm o}$ mode. 
In contrast, at the O(4)CP, the linear mixing
is banned by the underlying chiral symmetry.

\subsubsection{Susceptibility as a spectral sum}

Inverse of the differential operator (\ref{eq:normal})
with the
retarded boundary condition is
the response function
\begin{eqnarray}
\chi(\omega,q)=
\frac{\Gamma \lambda \bq^2}
{(-(\omega+i\varepsilon)^2 +\omega_{\rm o}^2)(-i\omega +|\omega_{\rm d}|)}
\left (
\begin{array}{cc}
\frac{-i\omega}{\lambda \bq^2}+1 &
-2\gamma \sigma \\
-2\gamma \sigma &
\frac{-\omega^2}{\Gamma}+\chi_h^{-1}+4\gamma^2\sigma^2 +\kappa \bq^2 
\end{array} \right ),
\end{eqnarray}
 which characterizes
the time--dependent response of these densities to the external fields,
$h$ and $j$,
within the linear approximation.
The susceptibility is obtained 
in the limit of $(\omega=0,\bq\to 0)$.
The response function is analytic in the upper
complex--$\omega$ plane, which fact allows us to express generally
the susceptibility as a sum of the mode spectra\cite{DF75}:
\begin{eqnarray}
\chi(0,\bq)=\frac{1}{\pi}\int \frac{d\omega}{\omega} 
{\rm Im} \chi(\omega,\bq),
\end{eqnarray}
where a ultra--violet regularization is understood if necessary.
This expression shows that the divergence at a critical point should come from
an infrared enhancement of the spectral function because 
the spectral function itself is usually integrable.

Using this expression we can examine the relative weight of
each mode contribution to the susceptibility.
In our case, the oscillating and diffusion modes
give spectral contributions
as
\begin{eqnarray}
\chi_h&=&\lim_{\bq\to0}
\frac{1}{\pi}\int \frac{d\omega}{\omega} {\rm Im}\chi_h(\omega,\bq)
=
\chi_h \left (\frac{ \chi_h^{-1} } { \chi_h^{-1}+4\gamma^2\sigma^2 }+
            \frac{4\gamma^2\sigma^2}{\chi_h^{-1}+4\gamma^2\sigma^2}
\right ),
\end{eqnarray}
and 
\begin{eqnarray}
\chi_j&=&\lim_{\bq \to0}
\frac{1}{\pi}
\int \frac{d\omega}{\omega} {\rm Im}\chi_j(\omega,\bq)
=
\chi_j \left  ( 0 + 1 \right ).
\end{eqnarray}
Here the first term in the bracket originates from the poles 
$\pm\omega_{\rm o}$ and the second from $\omega_{\rm d}$.

First, we note that only the diffusion--like
$\omega_{\rm d}$ pole contributes to the susceptibility of $\varphi$.
This is a robust result following from the conservation of 
the density $\varphi$.
Existence of the
current ${\bf j}$ such that 
$\partial _t \varphi +\nabla \cdot {\bf j}=0$
dictates that 
the frequencies of the modes contributing to
the $\varphi$ susceptibility must 
vanish as $\bq$ goes to zero. 
We can formally show that
the spectrum of the $\varphi$ response function 
behaves as
$\lim_{\bq \to 0}{\rm Im} \chi_j(\omega,\bq)/\omega 
\propto \delta(\omega)$ (see Appendix A).
Conversely,
we can state that softening of the
hydrodynamic mode must accompany the divergence of $\chi_j$.

Second, the ratio of the $\omega_{\rm d}$ spectral contribution
to the total in the scalar susceptibility,
\begin{equation}
R \equiv
 \frac{4 \gamma^2 \sigma^2}{\chi_h^{-1}+4 \gamma^2 \sigma^2}
=1-\chi_j^{-1},
\end{equation}
goes to unity at the TCP approached from the broken phase
and at the Z$_2$CP, 
which means that 
the leading divergence of the scalar susceptibility 
is also generated by the $\omega_{\rm d}$ spectrum
at these critical points.
Even at the O(4)CP approached from the broken
phase the $\omega_{\rm d}$ spectrum gives a finite portion of the
divergence $0<R<1$ since $\chi_h^{-1} \sim \sigma^2 \to 0$.
This result can be understood by rewriting the scalar response function as
\begin{eqnarray}
\chi_h(\omega, \bq) 
&=&
\chi_h^{(0)}(\omega,\bq)
{1  \over 1 -
 \Omega_{\sigma\varphi} \chi_j^{(0)}(\omega,\bq)
\Omega_{\sigma\varphi}\chi_h^{(0)}(\omega,\bq)  } ,
\end{eqnarray}
where $\chi^{(0)}_h(\omega,\bq)
=1/(L_\sigma(\omega+i\varepsilon)+\Omega_{\sigma\sigma})$
and
$\chi^{(0)}_j(\omega,\bq)
=1/(L_\varphi(\omega)+\Omega_{\varphi\varphi})$.
The denominator expresses the linear mixing between the ``bare'' 
$\tilde \sigma$ and
$\tilde \varphi$ modes through the coupling 
$\Omega_{\sigma\varphi} \propto \sigma$.
Even though the coupling becomes smaller as 
the O(4)CP is approached from the broken phase, softening of the
mediating ``bare'' $\tilde \sigma$ propagator provides $1/\sigma^2$ factor,
which results in the finite mixing of the $\omega_d$ mode in 
the scalar channel. This is a simple example 
indicating the importance of the mode coupling
near the critical point.

In summary, 
the $\tilde \sigma$ and $\tilde \varphi$ fluctuations
mix and form two kinds of eigenmodes,
$\omega_{\rm o}$ and $\omega_{\rm d}$.
We find that along the O(4) critical line
approached from the broken phase, 
the critical eigenmode
shifts from the sigma--meson like $\omega_{\rm o}$
to the diffusion like $\omega_{\rm d}$ mode at the TCP.
In contrast, when we approach the TCP from the symmetric phase, 
the scalar susceptibility $\chi_h$ is given completely
by the critical $\omega_{\rm o}$ spectrum without any mixing of
the $\tilde \varphi$ fluctuation.
At the Z$_2$CP
the $\omega_{\rm o}$ mode becomes a fast mode 
whereas the whole divergence comes from the critical softening
of the $\omega_{\rm d}$ spectrum with the hydrodynamic character.

\subsubsection{Cases with other types of motion}

In more microscopic NJL model calculation in the later section,
the mode with the hydrodynamic character is provided 
as the mode of landau-damping type,
contrary to the macroscopic analysis in the previous subsection,
and we may change the time evolution operator
accordingly as
$L_\varphi(\omega)=-i\omega/\lambda \sqrt{\bq^2}$.
More generally,
the sigma mode may be oscillating or relaxing
($L_\sigma(\omega)=-\omega^2/\Gamma$ or $-i\omega/\Gamma$) 
while the possible hydrodynamic fluctuation can be the
diffusion, landau-damping type, or sound motion
($L_\varphi(\omega)=
-i\omega/\lambda \bq^2,
-i\omega/\lambda \sqrt{\bq^2}$, or
$-\omega^2/\lambda \bq^2$).
Furthermore, couplings with other hydrodynamic modes, if exist,
are to be included to describe the correct dynamic behavior 
of the system\cite{HH77,SS04,KO04}. 
We stress here that our result on the
critical eigenmodes at the critical points
are independent on these ambiguities.
The only important fact is that the
operator $L_\varphi$ has the hydrodynamic character, and therefore
intrinsically soft mode of the system.
In any of these choices, we find the eigenmodes
for small $\bq^2$ as
\begin{eqnarray}
L_\sigma(\omega) &=& -  (\chi_h^{-1}+4 \gamma^2 \sigma^2 ),
\quad
L_\varphi(\omega)= -  \chi_j^{-1} .
\end{eqnarray}
Note that the typical mode frequencies of the
diffusion, landau-damping type and sound-like dispersions
vanish $\omega \to 0$ as $\bq \to 0$, and satisfy
the spectral property following from the conservation law
\begin{eqnarray}
\frac{1}{\pi}{\rm Im}\frac{1}{L_{\varphi}(\omega)+\chi^{-1}}
\to  \omega \delta (\omega)\chi
\quad \mbox{as} \ |\bq| \to 0  .
\end{eqnarray}
Finally we note that these hydrodynamic modes drops out
in the $\omega$ limit of the response function:
\begin{eqnarray}
\chi(0^+,\bzero)=\lim_{\omega \to 0}
\frac{1}{\pi}\int \frac{d\omega'}{\omega'-\omega-i\epsilon}
{\rm Im}\chi(\omega',\bzero)
=
\lim_{\omega \to 0}\frac{1}{\pi}\int \frac{d\omega'}{\omega'-\omega-i\epsilon}
\omega' \delta(\omega')\chi
=0.
\end{eqnarray}

\section{Nambu--Jona Lasinio model with a tricritical point}
\label{sec:3}

As an definite illustration,
we shall study the spectral contributions
of the collective modes 
at critical points in the NJL model,
and confirm that the result is consistent with the TDGL approach.
We remark here that, 
unlike in the TDGL approach,
there are no bare bosonic modes. 
The bosonic modes are dynamically generated through
the interaction between the quarks
and their softening causes the divergences at the critical points.

\subsection{Effective potential and susceptibilities}

We analyze the simplest version of the NJL model \cite{NJL,HK94,SPK92}
$
{\cal L} = \bar q  (i \!\!\! \not\! \partial -m) q +g [
(\bar q q)^2 + ( \bar q i \gamma_5 \tau^a q)^2]
$
in the mean field approximation
($\langle \bar q q \rangle = \sigma$=const, 
$\langle \bar q i \gamma_5 \tau^a q \rangle =\pi$=0).
The thermodynamics is described
by the effective potential\cite{LB96},
\begin{eqnarray}
\Omega(T,\mu,m;\sigma)/V&=&
-\nu \int {d^3 k\over (2\pi)^3}
[E - T \ln (1-n_+) -T \ln (1-n_-)]
+{1\over 4 g}(2g\sigma)^2,
\label{eq:NJLfree}
\end{eqnarray}
where $n_{\pm}=(e^{\beta(E\mp \mu)}+1)^{-1}$, $E=\sqrt{M^2+\bk^2}$,
$M=m-2g \sigma$, and $\nu=2 N_f N_c=2\cdot 2 \cdot 3=12$ with
$N_f$ and $N_c$ the numbers of flavor and color, respectively.
Here  $\mu$ is the quark chemical potential.
The true thermodynamic state is determined by the 
extremum condition, $\partial \Omega /\partial  \sigma=0$,
and the corresponding grand potential is $\Omega(T,\mu,m)$.
We define the model with the three--momentum cutoff 
$\Lambda$ and with the coupling constant $g \Lambda^2=2.5$
which allows the TCP.
In the following, all the dimensionful 
quantities are expressed in the units of $\Lambda$.

Expansion of the effective potential around $\sigma=0$ with
$m=0$ gives rise to
\begin{eqnarray}
\Omega(T,\mu,0;\sigma)/V&=&-\nu \int \frac{d^3k}{(2\pi)^3}
[k-T \ln(1-n^0_+) -T \ln(1-n^0_-)]
 +\frac{1}{2}(\frac{1}{2g}-J^0)(2g\sigma)^2
 +\frac{1}{2\cdot 4} I^0(2g\sigma)^4 + \cdots
\nonumber \\
&=&\Omega(T,\mu,0;0)/V
 + a(T,\mu)(2g\sigma)^2  + b(T,\mu)(2g\sigma)^4 + \cdots,
\label{eq:NJL0effpot}
\end{eqnarray}
where the superscript 0 indicates the quantity evaluated in the massless limit.
The first term is the non-singular part of the free energy 
in the GL description.
The integrals 
$I^0$ and $J^0$ are given in Appendix B.
The TCP determined by $a=b=0$
appears at
$T_t/\Lambda =0.20362$ and $\mu_t/\Lambda = 0.49558$. 

\begin{figure}[tb]
\begin{center}
\hfill
\epsfxsize=0.7\textwidth
\epsffile{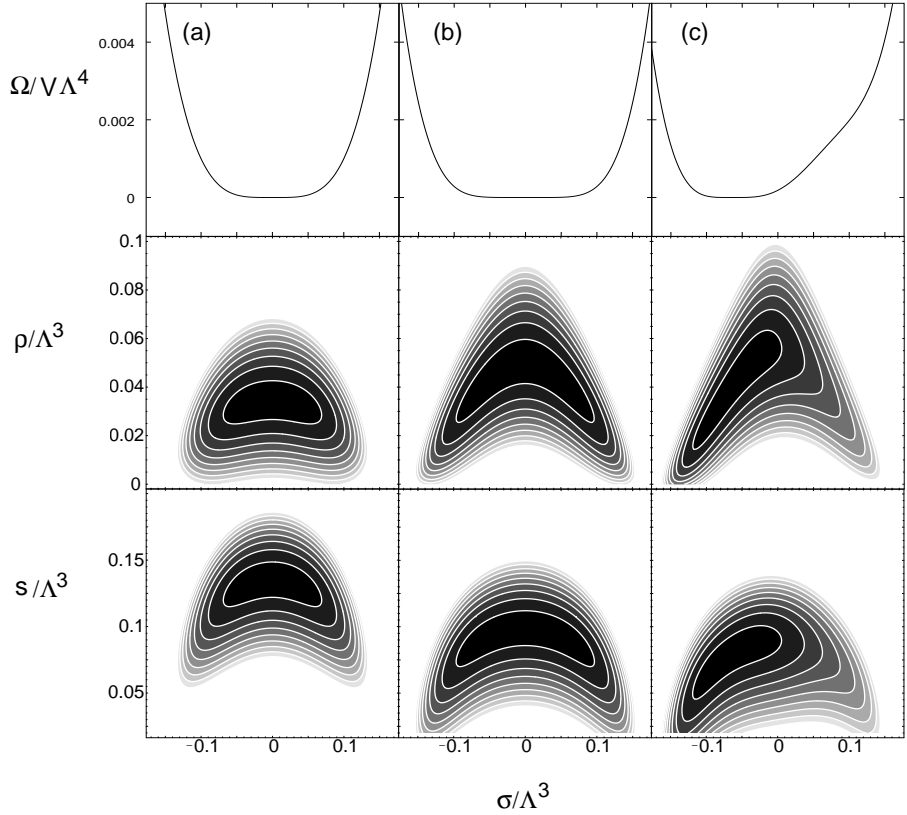}
\hfill
\vspace{2mm}
\caption{Effective potentials of the NJL model
at three critical points, (a) O(4)CP, (b) TCP and (c) Z$_2$CP.
In the upper panels shown are 
the potentials (\ref{eq:NJLfree}) 
measured from the minima
as functions of a single ordering density $\sigma$.
The middle and lower panels are the
contour plots of the potentials (\ref{eq:GL2})
with two ordering densities, ($\sigma, \rho$) and 
($\sigma, s$), respectively.}
\label{fig:GLpotential}
\end{center}
\end{figure}

As explained in \S II, 
it is useful to introduce 
the effective potential with another 
relevant ordering density besides the $\sigma$
in studying the behavior of the
quark number susceptibility and specific heat near
the TCP and the Z$_2$CP.
From the physical grand potential $\Omega(T,\mu,m)$,
we can construct the Landau effective potential with
two ordering densities  $\rho$ and $\sigma$ in the following way:
first we introduce the free energy $F(T,\rho,\sigma)$ via
\begin{equation}
F(T,\rho,\sigma)/V=\Omega(T,\bar \mu,\bar m)/V+\bar \mu \rho-\bar m \sigma,
\end{equation}
where $\bar \mu=\bar \mu(T,\rho,\sigma)$ and
$\bar m=\bar m(T,\rho,\sigma)$ are defined
by inverting the functions
\begin{eqnarray}
\rho&=&-\frac{1}{V}\frac{\partial \Omega}{\partial \mu}(T,\mu,m),
\quad
\sigma=\frac{1}{V}\frac{\partial \Omega}{\partial m}(T,\mu,m).
\end{eqnarray}
Then introducing new parameters
$\mu$ and $m$, we {\em define}
the Landau-type effective potential as
\begin{equation}
\tilde \Omega (T,\mu,m;\rho,\sigma)/V=
F(T,\rho,\sigma)/V - \mu \rho+m \sigma=
\Omega(T,\bar \mu,\bar m)/V + (\bar \mu - \mu) \rho
-(\bar m - m)\sigma.
\label{eq:GL2}
\end{equation}
The extremum condition for the densities $\rho$ and $\sigma$
yields $\mu=\bar \mu$ and $m=\bar m$, recovering
the physical grand potential $\Omega(T,\mu,m)$.
Use of the entropy density $s$ instead of the quark number
$\rho$ is straightforward.
It is known that the effective potential constructed in this way 
must be convex
and cannot be defined in the mixed phase.
Fortunately in the NJL model 
we can bypass this difficulty by 
supplementing the unphysical grand potential $\Omega(T,\mu,m)$ 
defined with the unstable solutions of the gap equation,
which corresponds to the non--convex part of the potential
$\tilde \Omega$.

One can easily show that the susceptibilities 
$\chi_{ij}=-{1 \over V}\partial ^2 \Omega/\partial i \partial j$
($i,j=T,\mu, m$)
are equal to the inverse of the curvature
matrix of the GL effective potential at the extremum point.
From this fact the divergence of the susceptibilities at 
a critical point is related to the appearance of 
a particular flat direction in the GL effective potential.

In Fig.~\ref{fig:GLpotential}
we show the effective potential
with two ordering densities at three critical points,
(a) O(4)CP with $(T,\mu)=(0.3419,0.3)$, (b) TCP,
and (c) Z$_2$CP with $(T,\mu,m)=(0.1498,0.5701,0.01)$ in the 
units of $\Lambda$.
The critical instability at these points is usually discussed
using the effective potential (\ref{eq:NJLfree})
with a single order parameter $\sigma$ based on chiral symmetry, 
which is shown in the upper panels.
The flat curvature of this potential means
the divergence of the scalar susceptibility.
At the O(4)CP
it is clear 
from the potential with two ordering
densities ($\sigma, \rho$), or ($\sigma, s$), 
that the $\sigma$ axis is indeed the symmetry direction of the system.
The densities $\rho$ and $s$ depend on $\sigma^2$ 
when calculated from (\ref{eq:NJLfree}) with $m=0$.
This fact is seen here as a quadratic 
bending of the potential valley.
Thus the fluctuation of these densities
are weaker
than that of $\sigma$, and
the susceptibilities of the quark number and the
entropy have the smaller exponent $\alpha$.

\begin{figure}[tb]
\begin{minipage}{0.67\textwidth}
\epsfxsize=\textwidth
\epsffile{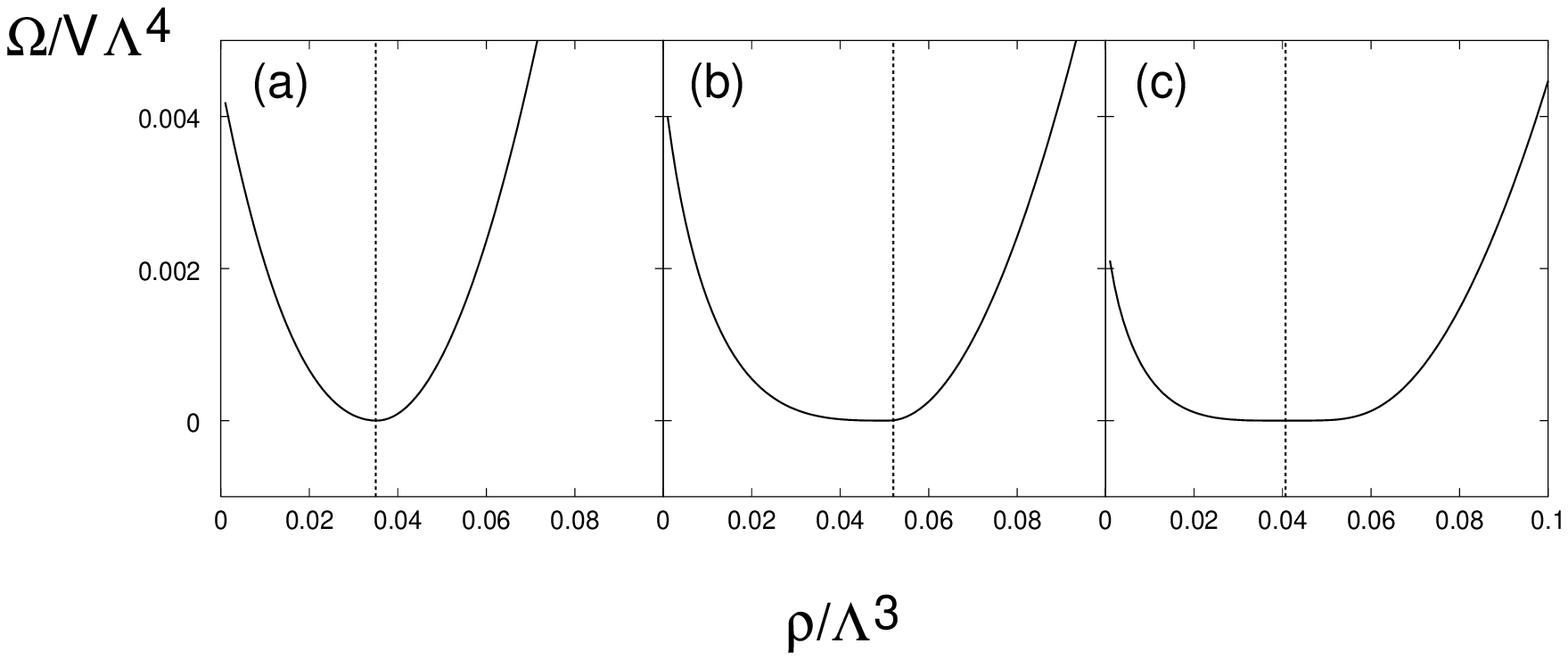}
\caption{Effective potentials of the NJL model as functions of a single ordering density $\rho$
at three critical points, (a) O(4)CP, (b) TCP and (c) Z$_2$CP. 
The dashed line indicates the critical density in each case.}
\label{fig:rhopot}
\end{minipage}
\hfill
\begin{minipage}{0.28\textwidth}
\epsfxsize=\textwidth
\epsffile{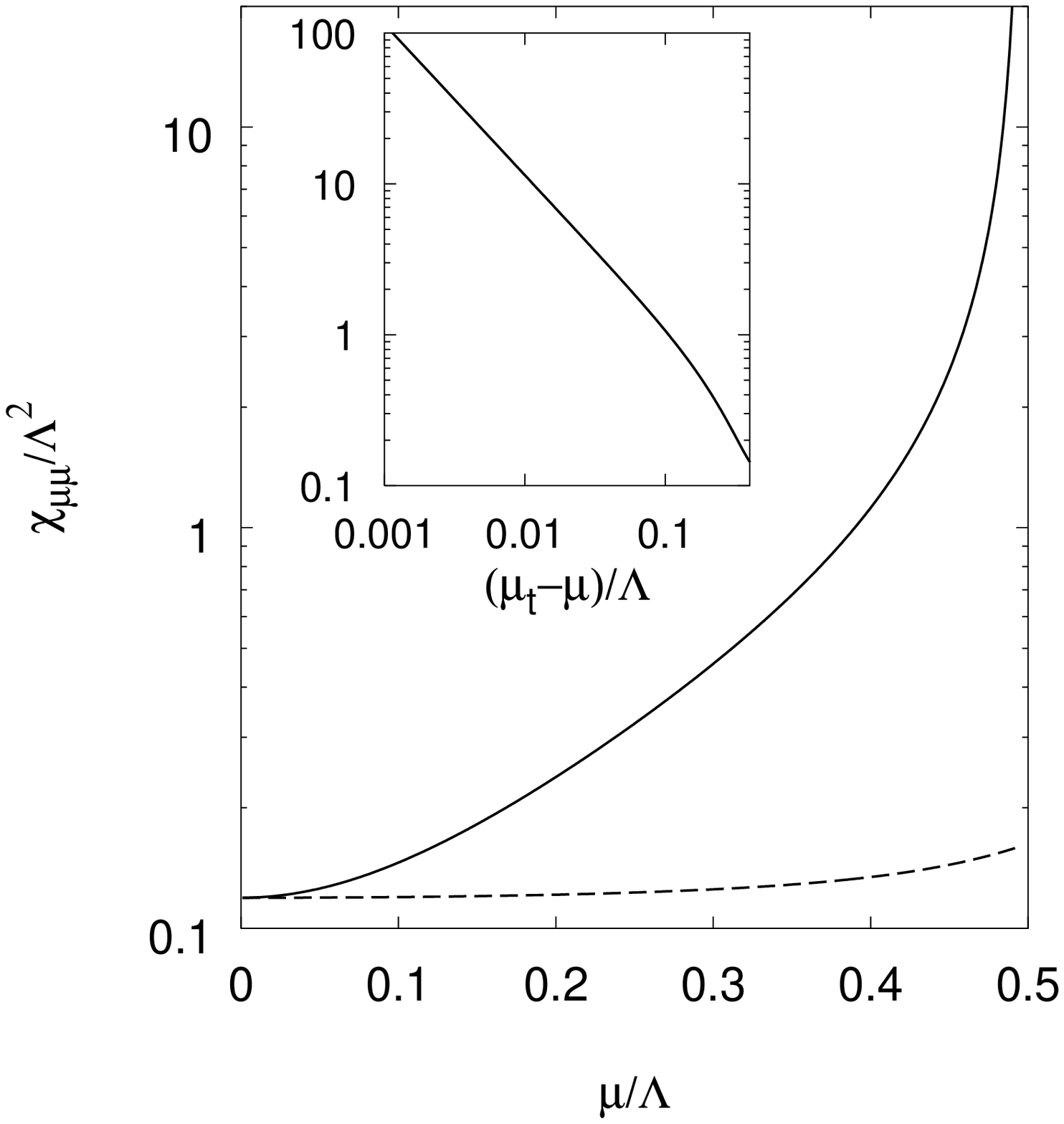}
\caption{$\chi_{\mu\mu}$ (solid line) along the O(4) critical
line approached from the broken phase. $\chi_{\mu\mu}$
approached from the symmetric phase is shown in a dashed line. 
Inset: $\chi_{\mu\mu}$ vs.\ $(\mu_t-\mu)/\Lambda$.
}
\label{fig:chimuAL}
\end{minipage}
\end{figure}

At the Z$_2$CP, on the other hand, the flat direction of the GL potential
is not parallel to the $\sigma$ axis 
in the $\sigma$--$\rho$ and $\sigma$--$s$ planes.
The proper flat direction is a {\em linear} combination of the
three densities of $\sigma$, $\rho$ and $s$,
and all the susceptibilities of them diverge
with the same exponent at the Z$_2$CP.

It will be very instructive to introduce the GL function with single
ordering density by eliminating $\sigma$ by 
$\partial \tilde \Omega/\partial \sigma=0$
in favor of $\rho$,
as shown in Fig.~\ref{fig:rhopot}. 
The curvature at the extremum coincides with
the inverse of the quark number susceptibility.
In case of the O(4)CP
the curvature does not vanish, implying the finite susceptibility
$\chi_{\mu\mu}$. It takes different values depending on from which
side we approach the equilibrium value of $\rho$.
Since the $\sigma^2$ and $\sigma^4$ terms of 
the potential (\ref{eq:NJL0effpot}) disappear at the TCP
and $\rho$ changes with $\sigma^2$ along the potential valley
(see Fig.~\ref{fig:GLpotential} (b)),
the $\rho$ potential becomes flat 
on the side corresponding to the broken phase as seen
in Fig.~\ref{fig:rhopot} (b).
This indicates the critical point for $\rho$.
On the higher density side, in contrast, the curvature is non--vanishing.
At the Z$_2$CP, the potential Fig.~\ref{fig:rhopot} (c)
is essentially the same as the potential 
(\ref{eq:NJLfree}), and
we may equally well  choose $\rho$ or $s$
as the ordering density instead of
$\sigma$ to describe this criticality.

We show in Fig.~\ref{fig:chimuAL}
the quark number susceptibility
$\chi_{\mu\mu}$ as a function of $\mu$ 
along the O(4) critical line,
across which $\chi_{\mu\mu}$ is discontinuous.
The value of $\chi_{\mu\mu}$ on the O(4) critical line
approached from the broken phase
grows up toward the TCP and eventually diverges there
as is described 
with the GL potential ($\chi_{\mu\mu} \propto 1/b$)\cite{HI03}.
The $\chi_{TT}$ also behaves in the same way.
The only qualitative difference is that 
at the point $\mu=0$
the $\chi_{\mu\mu}$ is continuous across the phase boundary
because no linear coupling with $\sigma$ is allowed
due to the symmetry under $\rho \leftrightarrow -\rho$.

\subsection{Response functions and mode spectra}
\label{subsec:2}

The spectral origin of the critical divergence can be investigated
by studying the spectral function which is obtained
as the imaginary part of the response function.
We discuss here the structure of the collective
eigenmode, which couples with the relevant susceptibilities
and shows softening  at the critical point.

The response functions in the NJL model
are calculated as\cite{HK94,SPK92,LB96}
\begin{eqnarray}
\chi_{ab}(iq_4,{\bf q})
&=&
\Pi_{ab}(iq_4,{\bf q})+\Pi_{a m}(iq_4,{\bf q})
{1 \over 1-2g \Pi_{m m}(iq_4,{\bf q})} 2g \Pi_{m b}(iq_4,{\bf q}),
\qquad ( a,b = \mu, m,\beta).
\label{eq:response}
\end{eqnarray}
Here the polarizations  are defined with the 
imaginary--time quark
propagator ${\cal S}(\tilde k)=1/(\not \tilde k + M)$ as
\begin{eqnarray}
\Pi_{ab}(iq_4,{\bf q})&=&
-\int {d^3 k \over (2 \pi)^3} T \sum_n 
tr_{\rm fcD} {\cal S}(\tilde k)\Gamma{\cal S}( \tilde k -q)
\Gamma',
\label{eq:polfuncs}
\end{eqnarray}
where $q_4=2l\pi T$ $ (l\in {\Bbb  Z})$, 
$\tilde k=({\bf k}, k_4+ i \mu)$, $\Gamma$ is an appropriate Dirac matrix,
and the trace is taken
over the flavor, color and Dirac indices.
$\Gamma=1$ for the scalar, $i\gamma_4$ for the baryon number,
and ${\cal H}_{MF}$ for $\beta$ with
\begin{eqnarray}
{\cal H}_{MF}&=&-i\frac{1}{2} \gamma_i \!
\stackrel{\leftrightarrow} {\nabla}_i + M  + i\mu  \gamma_4.
\end{eqnarray}
We calculate here the response function with ${\cal H}_{MF}$  the energy operator in the mean field
approximation instead of the entropy, 
because the entropy has no microscopic expression as it is defined
only in equilibrium.
The real--time response function is obtained 
from the imaginary--time propagator through the usual replacement
$iq_4 \to q_0 +i \epsilon$ in the final expression.
The static response function in the long wavelength limit
reduces to the corresponding susceptibility, 
$\lim_{\bq\to \bzero}\chi_{ab}(0,\bq)=\chi_{ab}$.
Especially, $\chi_{\beta\beta}=T^2 \chi_{TT}$.

The one--loop polarizations $\Pi_{ab}(\omega,\bq)$
are the response functions of the free quark gas with mass $M$,
and contain no contributions from the collective mode.
In our simple NJL model, the collective mode is generated by the
bubble sum encoded in the denominator of Eq.~(24)
and there are
two kinds of collective motion\cite{HF03}:
the sigma meson mode and the particle--hole (p--h) mode.
In Ref.~\cite{HF03} it is argued that the soft mode associated with the
Z$_2$CP is not the sigma meson but the p--h motion. 

\begin{figure}[tb]
\centerline{\hfill
\epsfxsize=0.45\textwidth\epsffile{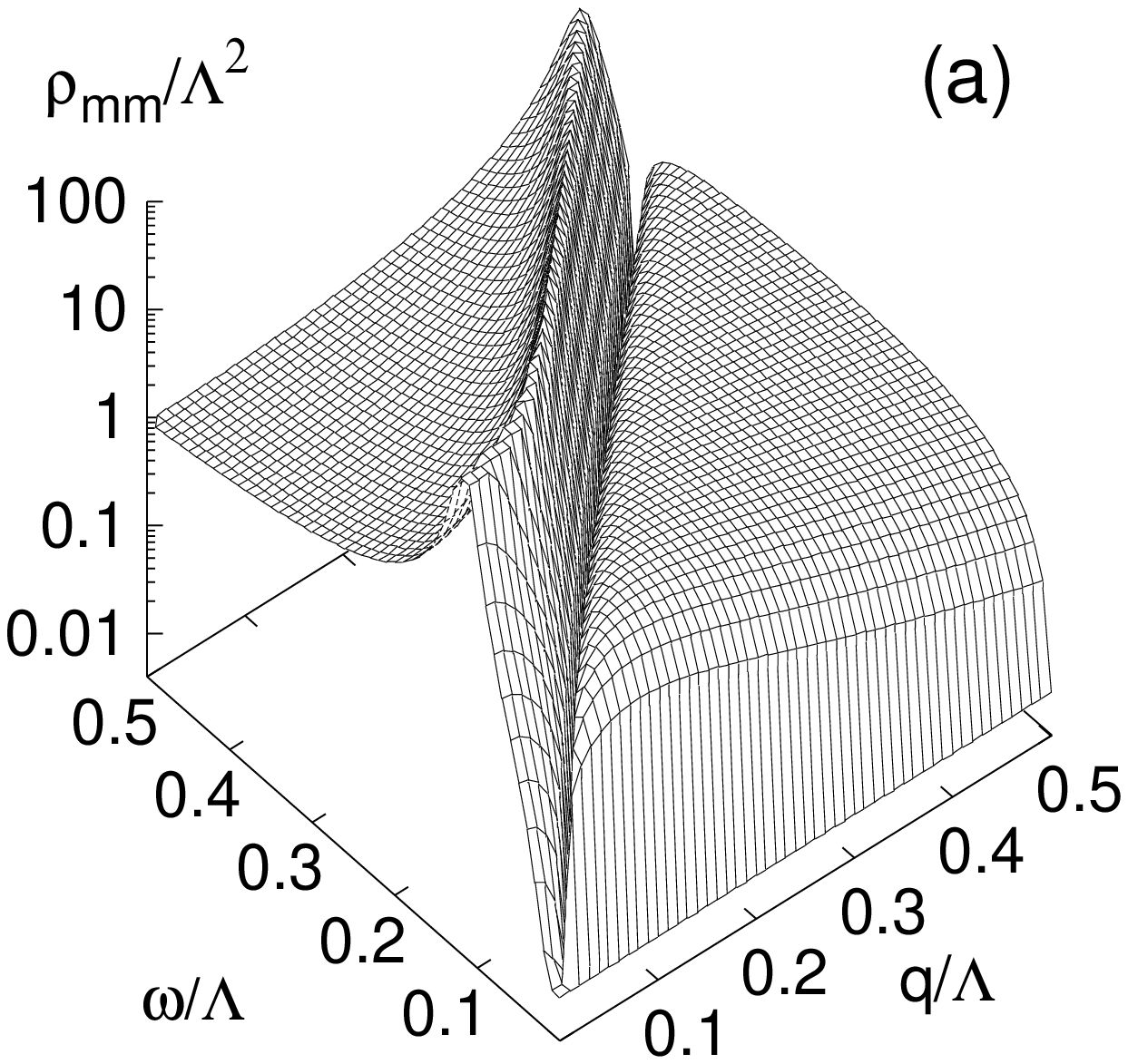}
\hfill
\epsfxsize=0.45\textwidth\epsffile{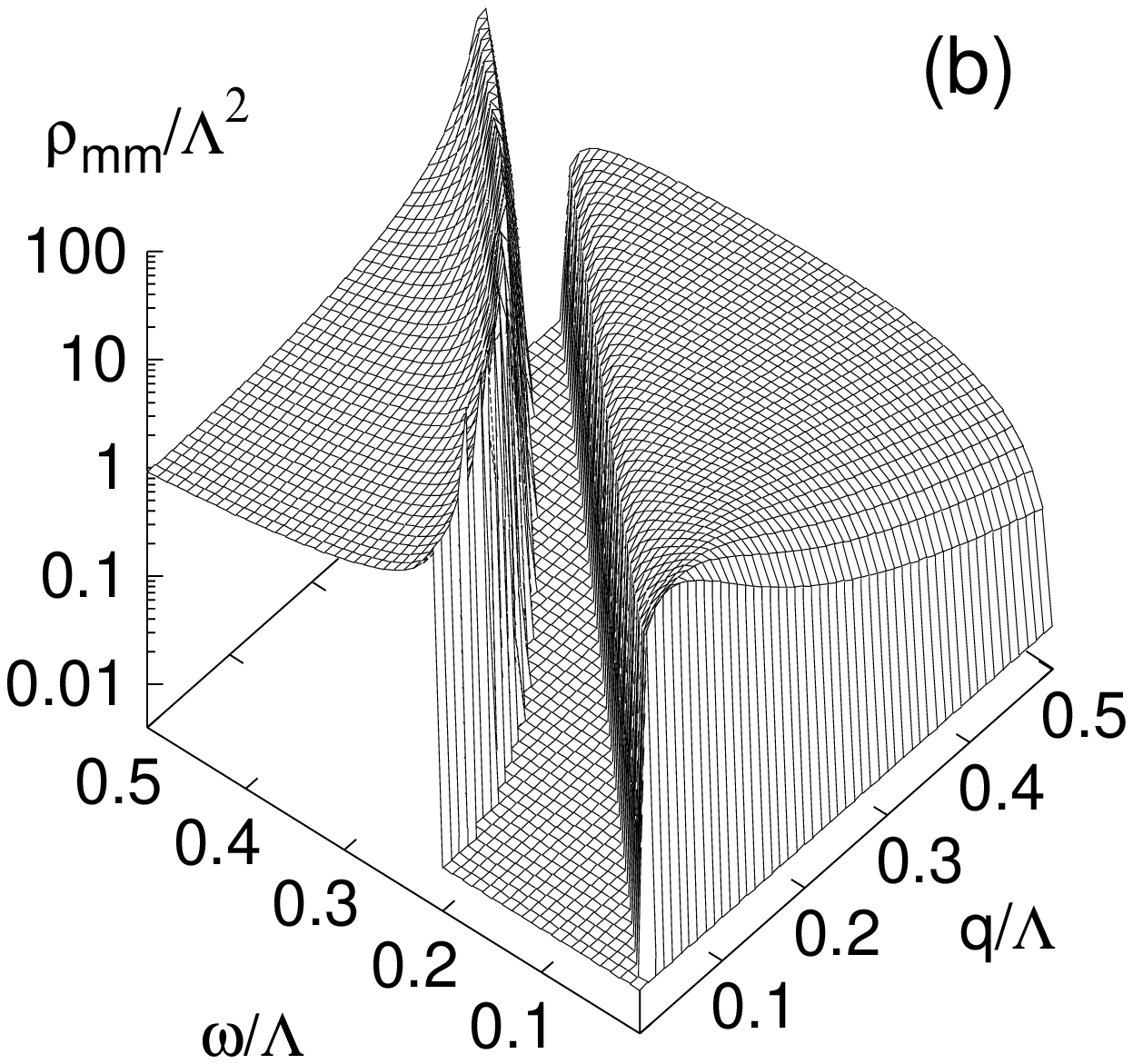}
\hfill}
\caption{Spectral functions of the scalar channel
in the $\omega$--$q$ plane
near the chiral critical point ($T_c/\Lambda,\mu_c/\Lambda)$=(0.3419, 0.3).
(a) $T/\Lambda=0.350$ and (b) $T/\Lambda=0.339$ with $\mu=\mu_c$ fixed.}
\label{fig:spfuncchiral}
\end{figure}

\begin{figure}[tb]
\centerline{\hfill
\epsfxsize=0.45\textwidth\epsffile{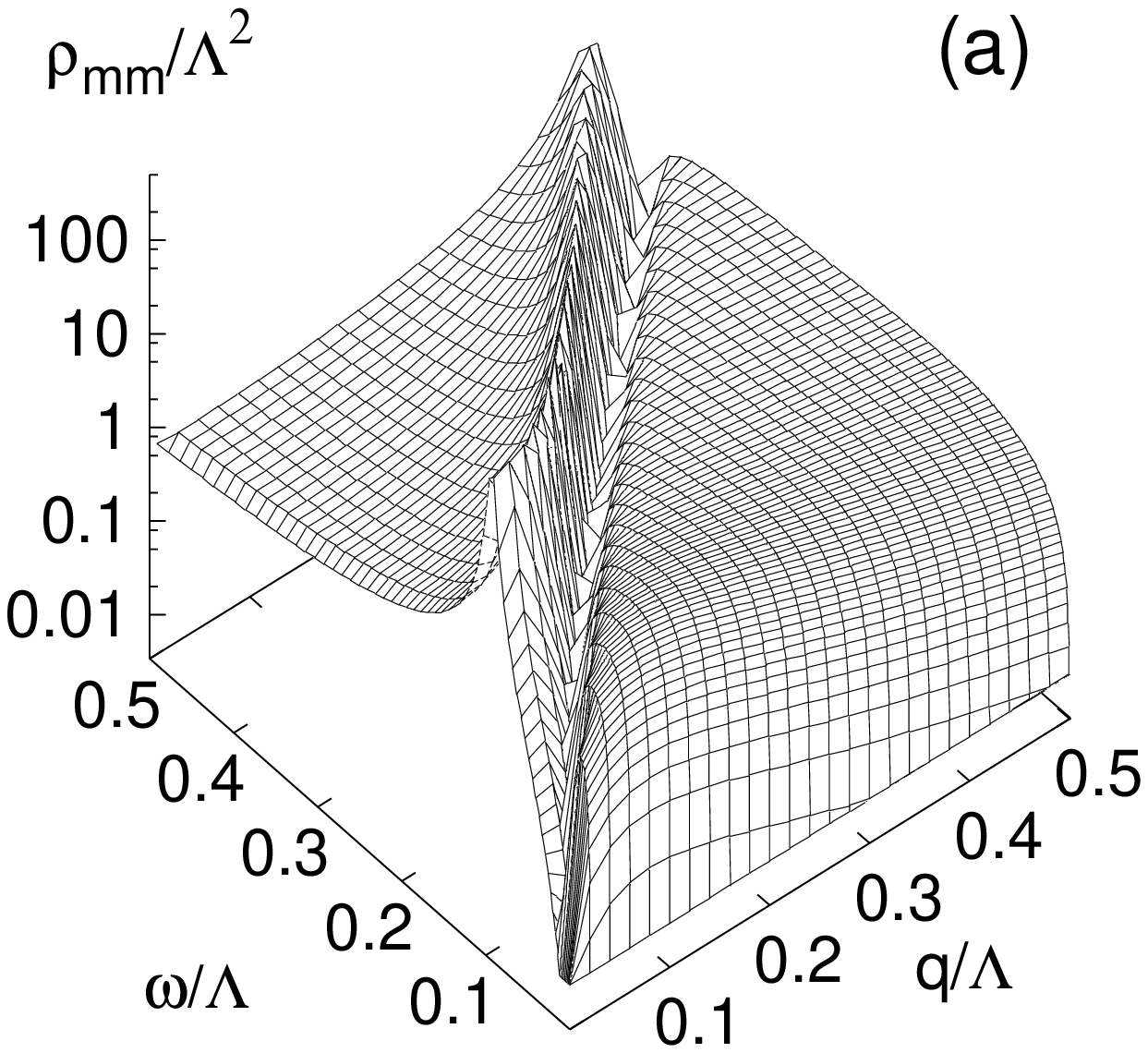}
\hfill
\epsfxsize=0.45\textwidth\epsffile{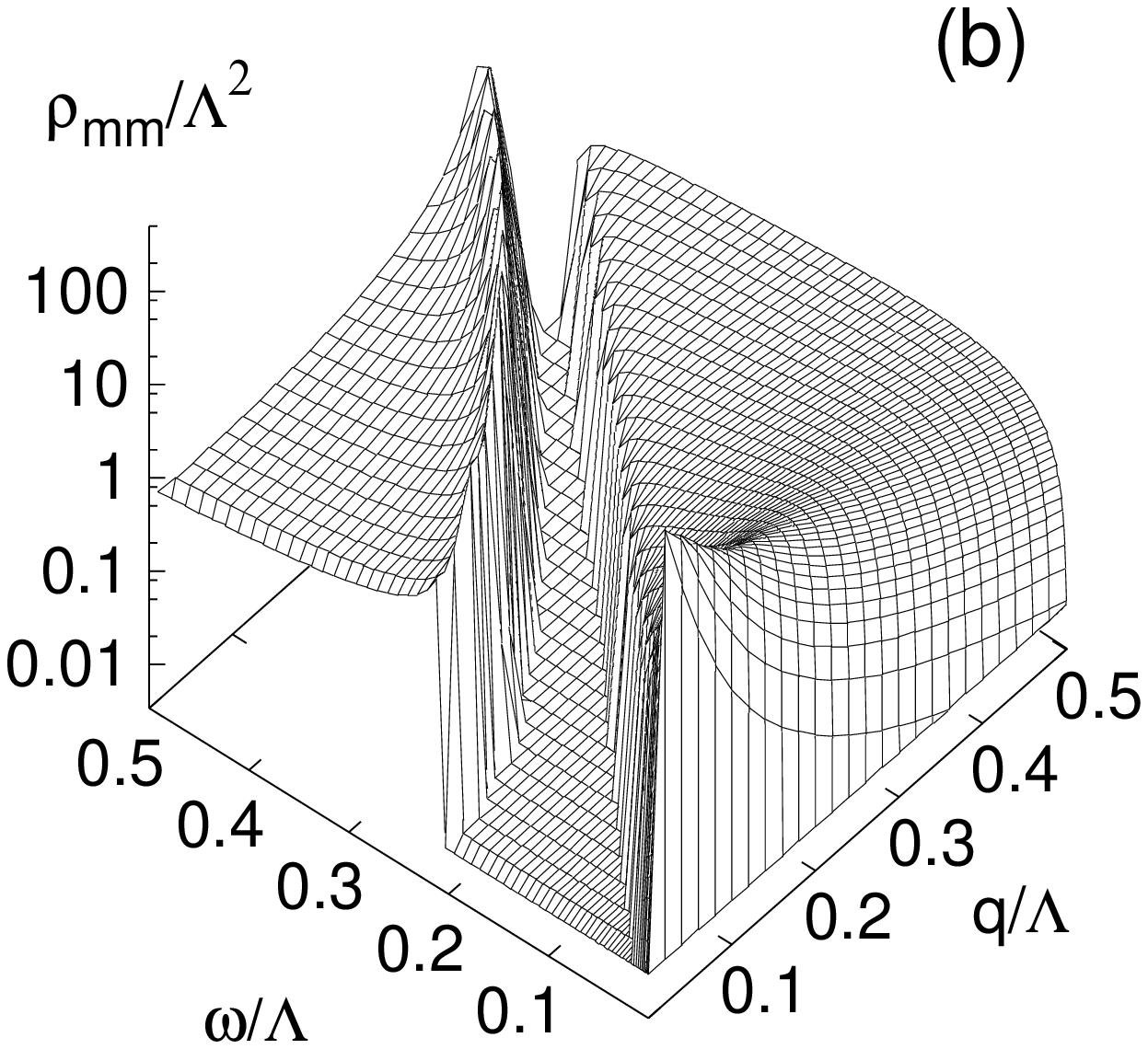}
\hfill}
\caption{Spectral functions of the scalar channel 
in the $\omega$--$q$ plane
near the TCP.
(a) $T/\Lambda=0.210$ and (b) $T/\Lambda=0.2035$ with $\mu=\mu_t$ fixed.}
\label{fig:spfuncTCP}
\end{figure}

\begin{figure}
\centerline{\hfill
\epsfxsize=0.47\textwidth\epsffile{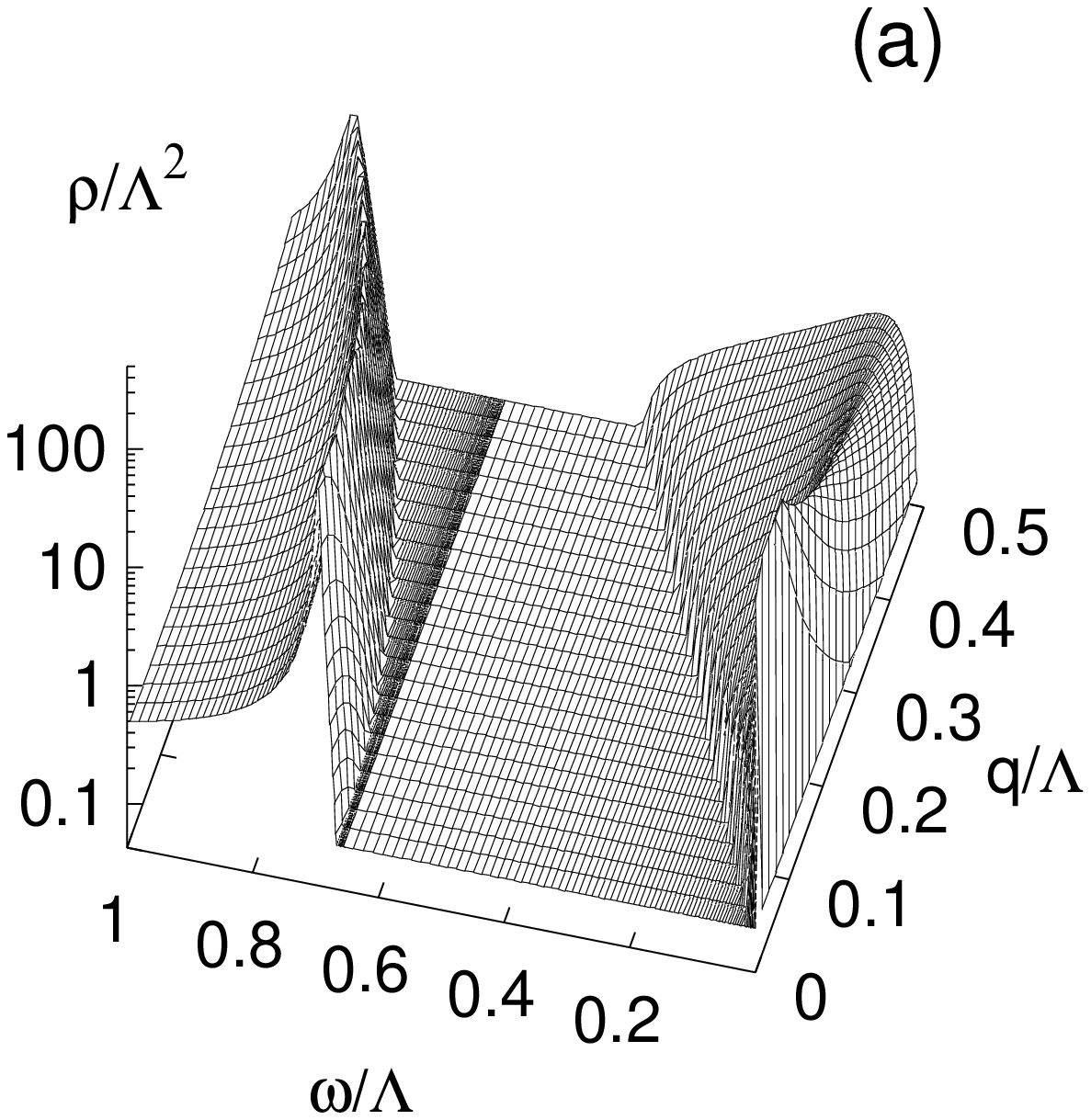}
\hfill
\epsfxsize=0.47\textwidth\epsffile{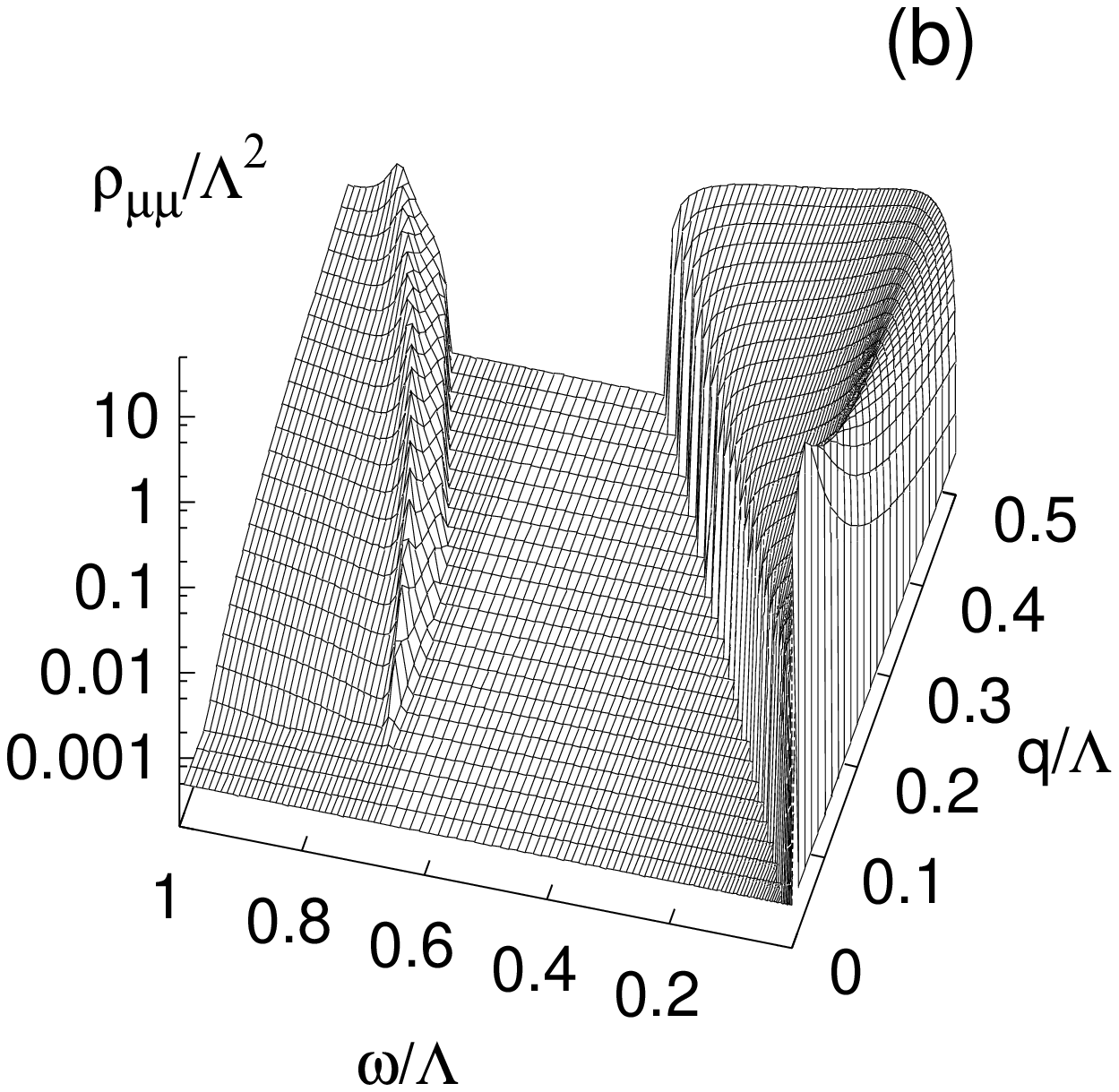}
\hfill}
\caption{Spectral functions of the scalar (a)
and quark number (b) response functions
at the Z$_2$CP with $m/\Lambda=0.01$.}
\label{fig:spfunc}
\end{figure}

The spectral function $\rho_{mm}(\omega,\bq)$
 of the scalar response function\cite{HK85} yields
\begin{eqnarray}
\rho_{mm}(\omega,\bq)
&=&2{\rm Im}\chi_{mm}(\omega, {\bf q})=
2{\rm Im}
{1 \over 2g} \left (
{1
\over 
1 -2g\Pi_{mm}(\omega,{\bf q})  }
-1 \right 
)\nonumber \\
&=&
{2{\rm Im} \Pi_{mm}(\omega, {\bf q}) \over 
[1- 2g {\rm Re}\Pi_{mm}(\omega,{\bf q})]^2+
[2g{\rm Im} \Pi_{mm}(\omega, {\bf q})]^2 } .
\label{eq:spectrum}
\end{eqnarray}
We notice that the 
spectrum $2{\rm Im}\Pi_{mm}(\omega, \bq)$
of the free quark gas is enhanced by the bubble--type correlation
in the denominator.

The scalar spectral functions 
are shown at $T/\Lambda$=0.350 and 0.339 near the O(4)CP
with $(T_c/\Lambda,\mu_c/\Lambda)$=(0.3419, 0.3)
in Fig.~\ref{fig:spfuncchiral}.
One should keep in mind
that the
$\delta(\omega -2M)$ spectrum of the sigma meson 
at $\bq=\bzero$ in the broken phase is hard to be seen
in this figure.
The sigma meson spectrum is softening
just above and below the O(4)CP (see also \S \ref{subsec:D}).
Besides the sigma spectrum
we clearly find the p--h mode spectrum in the space--like momentum region,
whose strength looks stronger in the broken phase.
As we approach the TCP,
as is shown in Fig.~\ref{fig:spfuncTCP},
this p--h mode spectrum grows
in the small $\bq$ region in the broken phase,
while in the symmetric phase it does not show such a enhancement.

We show in Fig.~\ref{fig:spfunc} the spectral functions of the scalar 
channel as well as the vector channel (quark number response)
at the Z$_2$CP with $m/\Lambda =0.01$. 
In the scalar channel 
clearly seen are the two spectral peaks 
of the sigma meson and the
p--h motion in medium, respectively\cite{HF03}.
This spectral structure is to be compared with
that of the free quark case given in Appendix B. 
The most significant feature in $\rho_{mm}$ is 
the critical enhancement in the $\omega\sim 0$ region
provided by the p--h mode, which gives rise to the divergence
of the scalar susceptibility.
Although the sigma meson shows the clear spectral peak
in this model, the mode is massive 
due to the explicit symmetry breaking by the current quark mass.

The spectral function of the quark number response,
$\rho_{\mu\mu}(\omega,\bq)$, also contains
these two spectral contributions, but
the sigma spectrum strongly diminishes as $\bq \to \bzero$.
It is worthwhile to note the fact that
  $\Pi_{m\mu}(\omega>0, \bzero)=0$ 
and ${\rm Im}\Pi_{m\mu}(\omega, \bzero)\propto \omega \delta (\omega)$
as $\bq \to \bzero$,
which reflects the conservation of the quark number.
Thus 
the response function $\chi_{\mu\mu}(\omega, \bq)$
obtained in (\ref{eq:response})
in the random phase approximation (RPA),
 too, shares the same property,
and 
the sigma meson cannot couple to the quark number susceptibility
$\chi_{\mu\mu}$ at $\bq=\bzero$.
Therefore the divergence of $\chi_{\mu\mu}$ at the Z$_2$CP must
comes solely from the softening of the p--h spectrum. The same is 
true for $\chi_{TT}$.

\subsection{Spectral sum along the critical line}

\begin{figure}[tb]
\begin{center}
\hfill
\epsfxsize=0.45\textwidth 
\epsffile{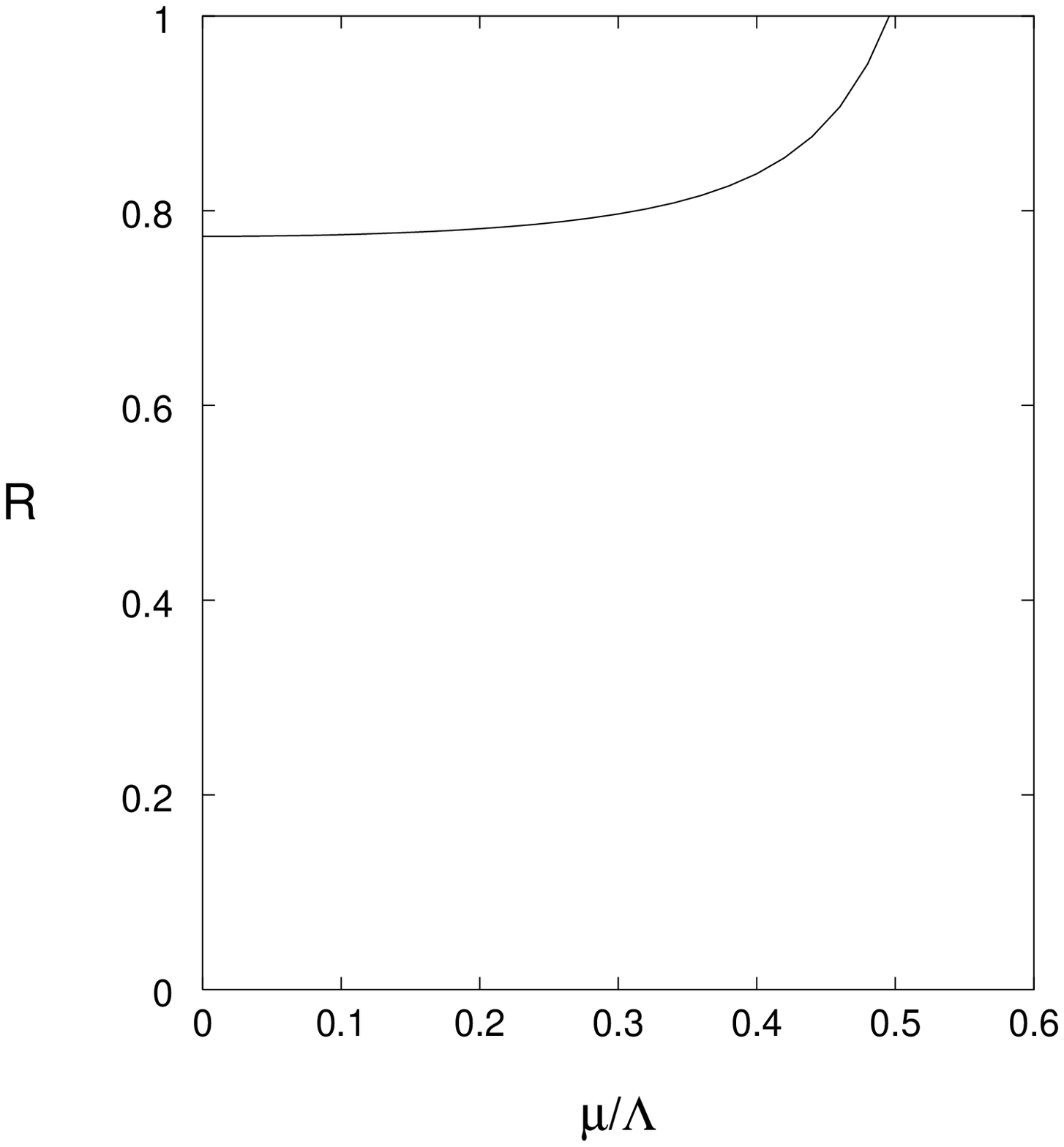}
\hfill
\vspace{5mm}
\caption{Ratio (\ref{eq:R-NJL}) of the spectral contribution along the chiral
critical line approached from the broken phase. $R \to 1$ toward
the TCP. }
\label{fig:ratio}
\end{center}
\end{figure}

In the previous subsection 
we have identified the soft mode associated with Z$_2$CP as the p--h
motion~\cite{HF03} generated in the scalar channel
whereas  at the O(4)CP 
the sigma meson mode becomes soft.
Let us examine how the changeover of the soft mode
from the sigma meson to the p--h one occurs along the critical line. 
Once we noticed that the difference between
the two limits, $\chi_{mm}(0^+,\bzero)$ and
$\chi_{mm}(0,\bzero^+)$, is caused by the 
hydrodynamic mode spectrum $\omega \delta (\omega)$ as $\bq \to \bzero$,
it is easy to calculate the ratio of the spectral strength of the two 
types of spectral contributions.
In these limits the explicit form of the RPA scalar response
functions in the broken phase (with $m=0$) yields, respectively,
\begin{eqnarray}
\chi_{mm}(0^+, \bzero) &=&
 \frac{1}{2g}\left (\frac{1}{2g \cdot 4 M^2 I(0^+,\bzero)}-1 \right ),
\nonumber \\
\chi_{mm}(0, \bzero^+) &=&
 \frac{1}{2g}\left (\frac{1}{2g \cdot 4 M^2 I(0,\bzero^+)}-1 \right ),
\end{eqnarray}
where function $I(\omega,\bq)$ is given in Appendix B.
Then we can define the ratio $R$ of the hydrodynamic spectrum to
the total strength of the scalar susceptibility
in the NJL model as
\begin{eqnarray}
R
&\equiv&\frac{\chi_{mm}(0,\bzero^+)-\chi_{mm}(0^+,\bzero)}
             {\chi_{mm}(0,\bzero^+)}
=\frac{I(0^+,\bzero)-I(0,\bzero^+)}{I(0^+,\bzero)}.
\label{eq:R-NJL}
\end{eqnarray}
On the other hand, these limits 
in the symmetric phase result in the same value
\begin{eqnarray}
\chi_{mm}(0, \bzero) &=&
 \frac{1}{2g}\left (\frac{1}{1-2g J^0}-1 \right ),
\end{eqnarray}
which means no hydrodynamic contribution to
the scalar susceptibility there ($R=0$).
The p--h mode must be decoupled from $\chi_{mm}$ 
in the symmetric phase.
Meanwhile we know that
the corresponding ratios for the susceptibilities of the conserved quantities
are always unity ($R=1$), which can be explicitly seen with the expressions
given in Appendix B.

The ratio $R$ (\ref{eq:R-NJL})
is shown in Fig.~\ref{fig:ratio}  as a function of
$\mu$ along the critical line.
We find that even in the O(4) chiral transition at zero
baryon number density ($\mu=0$)
the hydrodynamic spectrum contributes to the divergence
by a finite fraction.
This contribution of the
hydrodynamic spectrum increases toward the TCP,
and eventually gives the leading divergence
at the TCP, where $I(0,\bzero^+)=I^0=0$ but $I(\bzero^+,0) \ne 0$.
This behavior is completely in parallel with the TDGL approach.

The fact that the p--h mode gives a finite fraction
of the divergence at the O(4)CP
might be again unexpected from the viewpoint of
the sigma meson as the associated soft mode there.
Indeed, the sigma meson spectrum generates the
total divergence when the critical point is approached from the
symmetric phase. 
We should note here that  
the mixing of the scalar fluctuation in the broken phase
is the origin of the discontinuity
of the baryon number and energy susceptibilities
across the boundary and that 
only the scalar p--h mode with the hydrodynamic character 
can couple with the fluctuations of these conserved quantities.
Since the transitions between the
scalar and other channels is proportional to $M$,
the p--h spectral strength in the scalar is
necessarily of order $1/M^2$ so as to bring
a finite contribution to $\chi_{\mu\mu}$ and $\chi_{TT}$.

We note that 
the {\em scalar} p--h motion of the NJL model   
is possible only in medium, but
always possible in medium even in the symmetric phase,
where the p--h contribution should be decoupled from
the scalar susceptibility.
One may ask the reason for this decoupling of the p--h spectrum.
The absorption amplitude of the collective p--h 
mode with  momentum $\bq$ by a left--handed quark
$q_L(\bk)$ is proportional to a spinor product
$\bar u_R(\bk+\bq)u_L(\bk)$.
In the symmetric phase this coupling must contain
a helicity flip because 
chirality and helicity are identical for a massless quark,
and vanishes in the $\bq \to 0$ limit.
It is easy to show that $\bar u_R(\bk+\bq)u_L(\bk)\propto |\bq|$.
For a massive quark in the broken phase,
each of the left-- and right--handed states has
both helicity components, and thus 
$\bar u_R(\bk+\bq)u_L(\bk)\to 2M \ne 0$ in the $\bq \to 0$ limit.

\subsection{Behavior of poles and residues}
\label{subsec:D}

Let us discuss a little more details of the spectral contributions 
to the susceptibility, studying the
the poles  and the residues of
the scalar response function 
\begin{eqnarray}
\chi_{mm}(\omega,\bq)=
{1 \over 2g} \left ( {1 \over 1-2g J(\bq)+2g (4M^2-q^2)I(\omega,\bq)}
-1\right )
\end{eqnarray}
near the NJL  critical points.
It is useful to represent the spectral contributions as
\begin{equation}
\chi_{mm}^{\rm pole}(\omega,\bq) =  \sum_{i=\pm \sigma,{\rm ph}}
{R_i(\bq)  \over -\omega +\omega_i(\bq)}
\end{equation}
with the poles corresponding to the sigma meson ($\pm \sigma$)
near $\pm 2M$ and the p--h (ph) mode on the negative imaginary axis\cite{HF03}.
We would obtain the susceptibility as $\bq\to \bzero$ after
setting $\omega = 0$. 
As for the p--h contribution, however, we take into
account the kinematic condition $|\omega/\bq| <1 $ for the spectrum
via
\begin{equation}
{1 \over 2\pi}
\int_{-|\bq|}^{|\bq|} {d\omega \over \omega}
2{\rm Im}
{R_{\rm ph}(\bq)  \over -\omega +\omega_{\rm ph}(\bq)}
= 
{R_{\rm ph} \over \omega_{\rm ph}}\cdot 
{2 \over \pi} \tan^{-1}{|\bq| \over | \omega_{\rm ph}|}.
\end{equation}
The scalar susceptibility in this approximation is expressed
as a sum of the sigma and p--h pole contributions:
\begin{equation}
\chi_{mm}^{\rm pole}(0,\bzero^+)=
2 {R_\sigma \over \omega_\sigma}+
\lim _{\bq \to \bzero}
{R_{\rm ph} \over \omega_{\rm ph}}\cdot 
{2 \over \pi} \tan^{-1}{|\bq| \over | \omega_{\rm ph}|}.
\end{equation}

In Fig.~9 we show numerical results of the poles and residues
of the scalar response function (32) as functions
of $t=|T-T_c|/T_c$ with fixed $\mu=\mu_c$.
The behavior of them can be understood
as follows:

{\em Across the O(4)CP}
\quad
In the broken phase ($1-2g J(\bzero)=0$),
the sigma pole with $\bq=0$ locates
at $\omega_\sigma=2M$ on the real axis,
whose residue is
$R_\sigma \sim 1/(M I(2M,\bzero))$.
These quantities  scale 
as $1/R_\sigma \sim M \sim \chi_{mm}^{-1/2} \sim t^{1/2}$.
In the symmetric phase ($M=0$),
$\chi_{mm}^{-1} \sim 1-2g J(\bzero) \sim t$.
The complex sigma meson pole appears
at $\omega_\sigma \sim \sqrt{\chi_{mm}^{-1}/I(\omega_\sigma,\bzero)}
\sim t^{1/2}$ 
with the
residue $\sim 1/\sqrt{\chi_{mm}^{-1} I(\omega_\sigma,\bzero)}
\sim t^{-1/2}$.
In both cases, the sigma mode gives
the appropriate strength of the divergence 
$R/\omega_\sigma \sim \chi_{mm} \sim t^{-1}$.

The p--h mode arises from the $\omega$ dependence of 
the function $I(\omega,\bq)$. 
Since we are interested in the behavior in the small $\omega$
and $\bq$ region
with $u\equiv \omega /|\bq|<1$,
we may approximate the function $I$ as
\begin{equation}
I(\omega ,\bq)=I(0,\bzero^+) + i u {\rm Im} I(0)' ,
\end{equation}
where 
${\rm Im}I(u)'=(d/du){\rm Im}I(u|\bq|,\bq)|_{\bq=\bzero}$.
In the broken phase, 
the condition, $I(\omega, \bq)=0$ gives
the pole as $\omega_{\rm ph} \sim -i |\bq|I(0,\bzero^+)$,
whose residue $R_{\rm ph}\sim -i |\bq|/M^2 \sim t^{-1}|\bq|$.
Similarly in the symmetric phase the condition
$1/(2g)-J(\bzero) - \omega^3/|\bq| \cdot i {\rm Im}I(0)'=0$
fixes the pole position as
$\omega_{\rm ph} \sim -i (\chi_{mm}^{-1} |\bq|)^{1/3} \sim (t|\bq|)^{1/3}$
with the residue
$R_{\rm ph} \sim -i |\bq|/
(\chi_{mm}^{-1} |\bq|)^{2/3} \sim t^{-2/3}|\bq|^{1/3}$. 
Then according to Eq.~(34) the spectral contribution is estimated to be
\begin{eqnarray}
&&
{ R_{\rm ph} \over \omega_{\rm ph}}
\tan^{-1}{ |\bq| \over |\omega_{\rm ph}|}
\sim {1 \over t},
\\
&&
{ R_{\rm ph} \over \omega_{\rm ph}}
\tan^{-1}{ |\bq| \over \omega_{\rm ph}}
\sim {1 \over t}\tan^{-1}{|\bq|^{2/3} \over t^{1/3}} \to 0 
\end{eqnarray}
for the broken and symmetric cases, respectively, as $|\bq| \to 0$.
We note that the p--h mode gives finite portion of 
the divergence at the O(4)CP
approached from the broken phase due to the
enhancement of the residue by $1/M^2$, despite that
the frequency $\omega_{\rm ph} \sim |\bq|$ shows no critical 
slowing.
The decoupling of the p--h mode 
in the symmetric phase is correctly described
by the behavior of the pole.

\begin{figure}[tb]
\centerline{\hfill
\epsfxsize=0.3\textwidth\epsffile{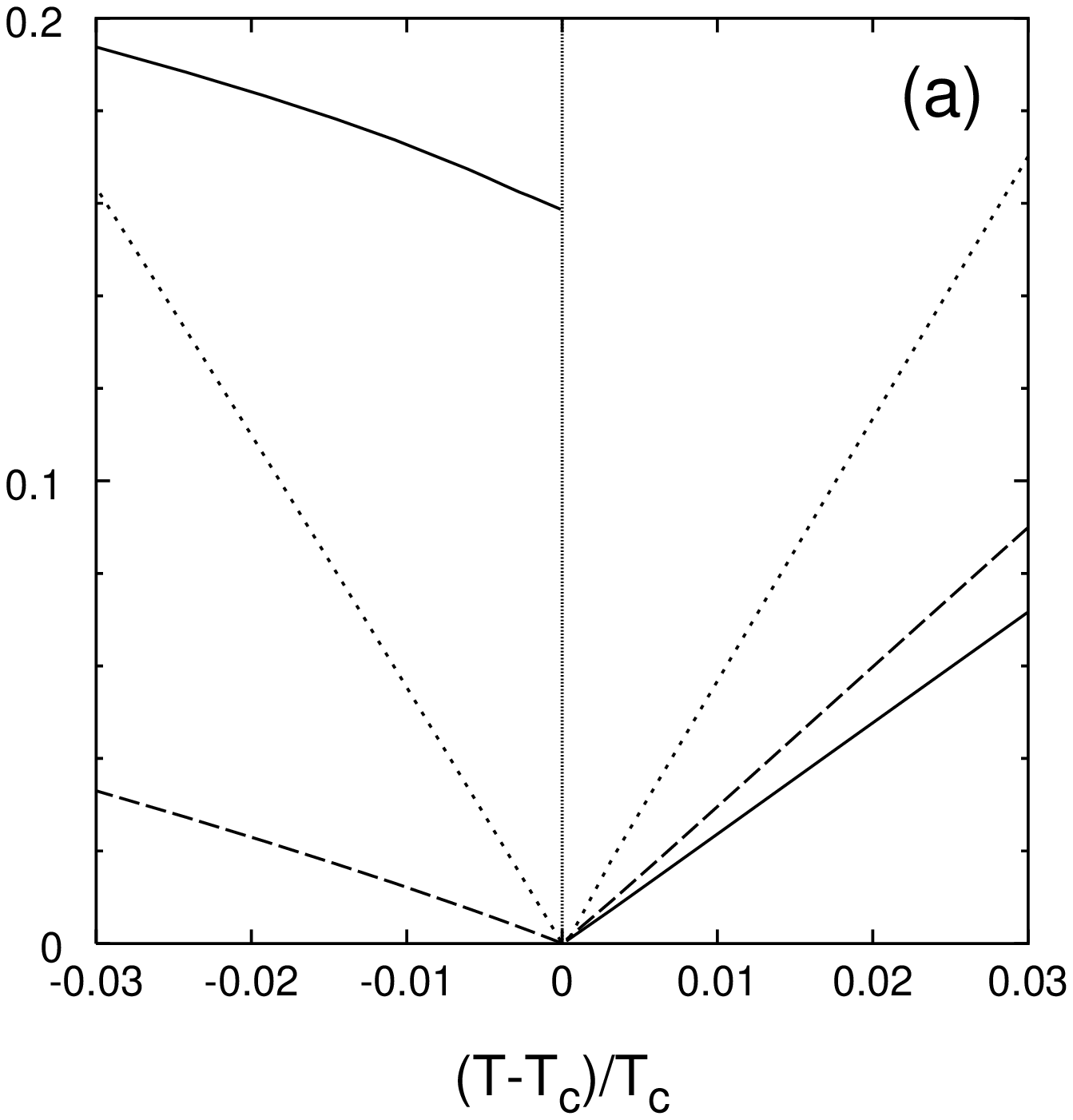}
\hfill
\epsfxsize=0.3\textwidth\epsffile{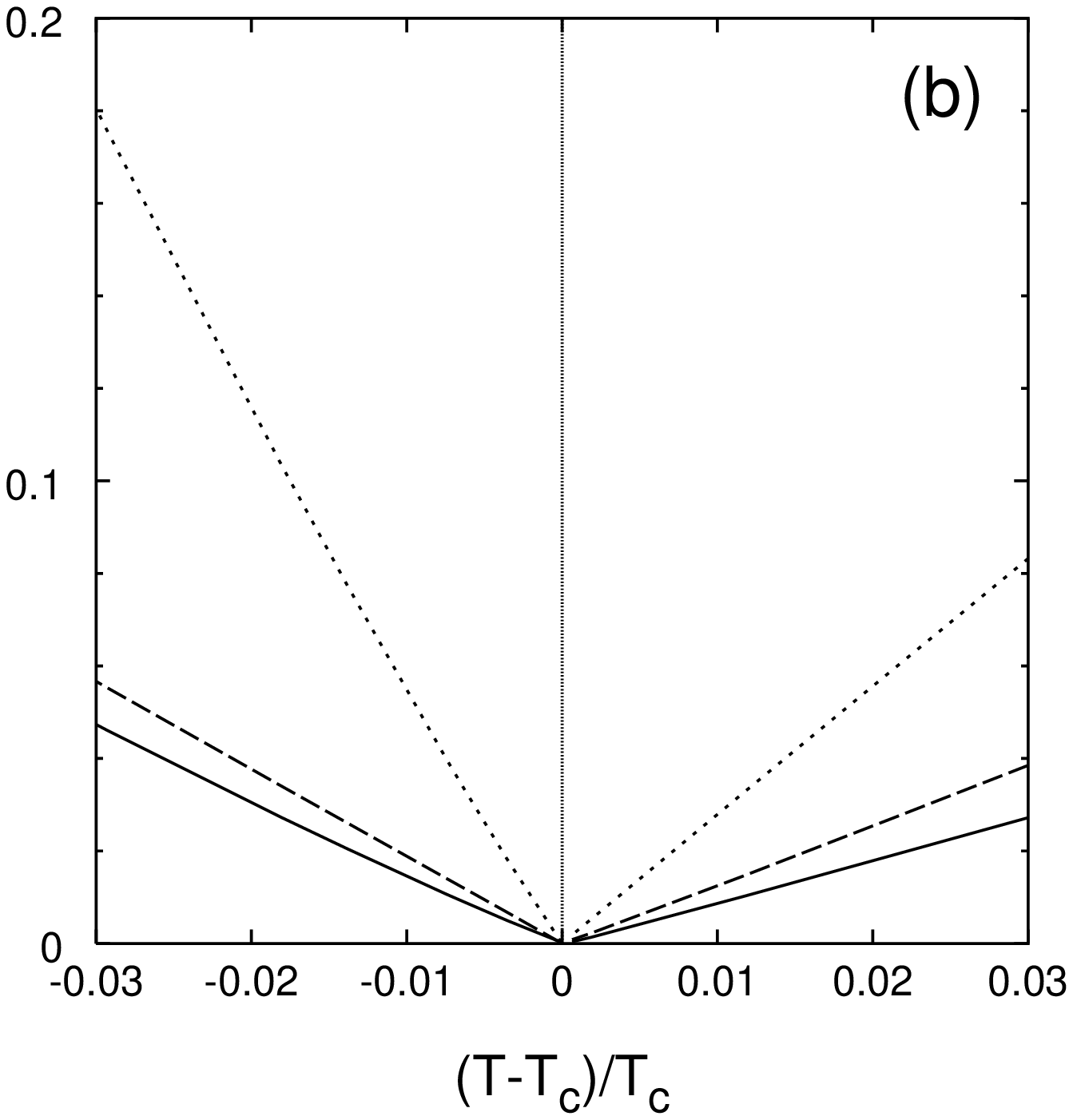}
\hfill
\epsfxsize=0.3\textwidth\epsffile{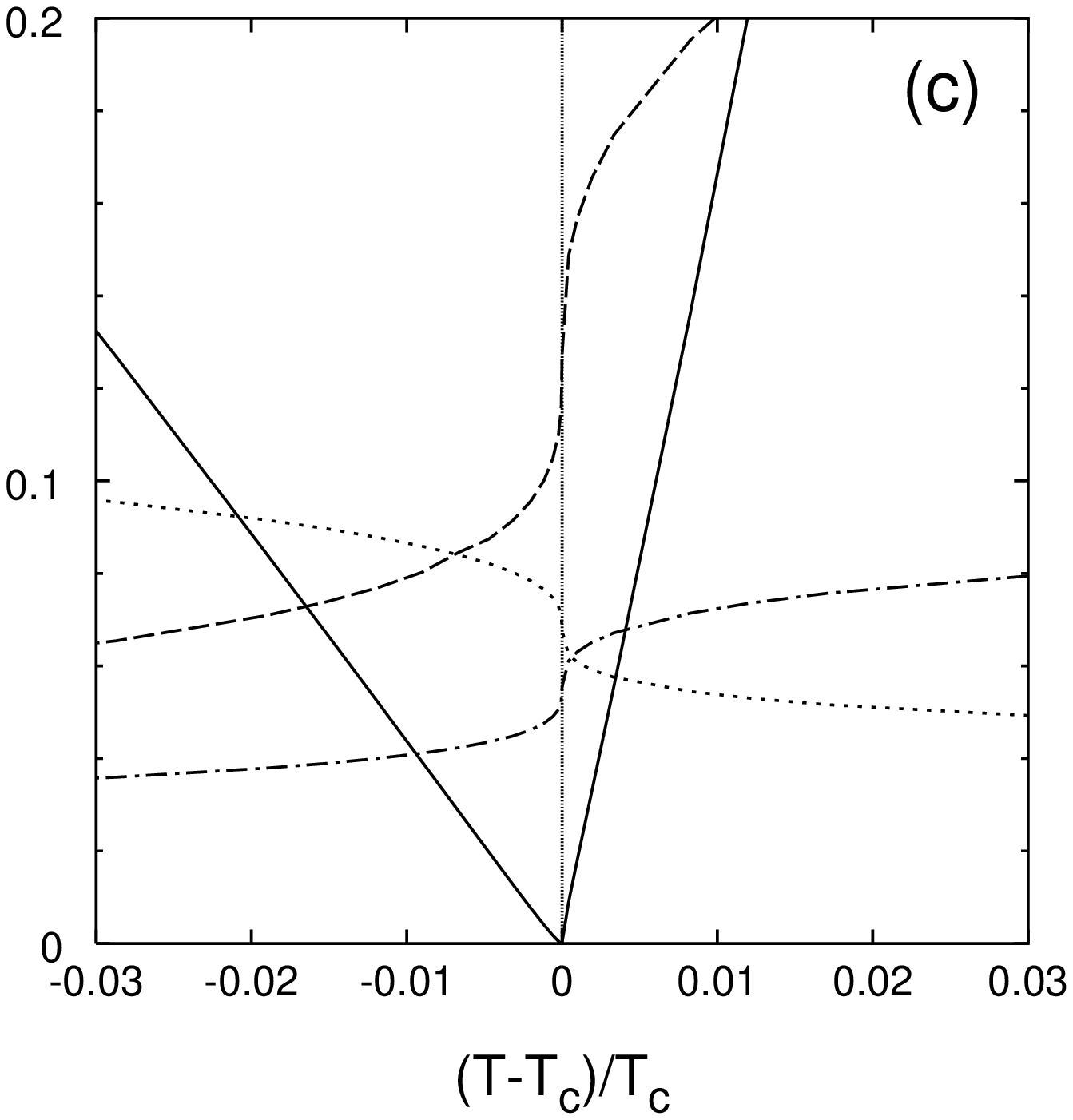}}
\vspace{2mm}
\caption{Scaling of the pole positions and 
the residues near the critical points. 
(a) O(4)CP: 
The solid line denotes 
$|\omega_{\rm ph}|/|\bq|$ for $t<0$ and 
$|\omega_{\rm ph}|^3/|\bq|$ for $t>0$ with $\bq=0.01$.
The residue $|R_{\rm ph}|^{-2}$ for $t<0$ and
$|R_{\rm ph}|^{-2/3}$ for $t>0$ is shown in a dashed line.
The sigma pole $\omega_\sigma^2=4M^2$ with
$\bq=\bzero$ for
$t<0$ and  $({\rm Re}\omega_\sigma)^2$ for $t>0$
in a dotted line.
(b) TCP: 
The similar plot to the O(4) case, but 
$(|\omega_{\rm ph}|/|\bq|)^2$ and
$\omega_\sigma^4$ for $t<0$.
(c) Z$_2$CP:
$(|\omega_{\rm ph}|/|\bq|)^{3/2}\times 10$,
$|R_{\rm ph}|$ and 
${\rm Re}\omega_\sigma$ are shown 
 in solid, dashed and dotted lines, respectively.
The $|{\rm Re} R_{\sigma}|$ is also drawn in a dash--dotted line.
}
\label{fig:polesresidues}
\end{figure}

{\em Across the TCP}\quad
At the TCP, $I(0^+,\bzero) \ne I(0,\bzero^+)=I^0=0$.
When we approach from the broken phase
$\chi_{mm}^{-1}=(2g)^2 4M^2 I(0,\bzero^+)\sim t$
while $\omega_\sigma \sim M \sim t^{1/4}$.
Then the sigma mode slows down, but cannot generate the 
leading divergence because
$R_\sigma/\omega_\sigma \sim t^{-1/2}$ whereas
$\chi_{mm} \sim t^{-1}$. 
The fact
$I(0,\bzero^+)\sim t^{1/2}$ 
changes the scaling of the p--h mode into
$\omega_{\rm ph} \sim t^{1/2} |\bq|$
with the residue $t^{-1/2}|\bq|$, which gives rise to
the correct order of the divergence
\begin{eqnarray}
&&
{ |R_{\rm ph} | \over |\omega_{\rm ph}|}
\tan^{-1}{ |\bq| \over \omega_{\rm ph}}
\sim 
{ 1 \over t }
\tan^{-1}{ 1 \over t^{1/2}}
\sim {1 \over t} .
\end{eqnarray}
The p--h mode must correspond to the critical eigenmode.
On the other hand, if the TCP is approached from the symmetric
phase, the p--h mode is decoupled from
the scalar susceptibility
and the critical sigma  mode generates the total
divergence.
It is very interesting that the soft mode associated with the TCP is
different between the symmetric and broken phases.

{\em At the Z$_2$CP}\quad
The sigma meson mode has
a finite energy gap of order $2M$ in our model. 
The pole position of the p--h mode can be evaluated as
$\omega_{\rm ph} \sim -i \chi_{mm}^{-1} |\bq| \sim t^{2/3}|\bq|$
with its residue $\sim -i |\bq| $, which
gives rise to
\begin{eqnarray}
&&
{ |R_{\rm ph} | \over |\omega_{\rm ph}|}
\tan^{-1}{ |\bq| \over \omega_{\rm ph}}
\sim \chi_{mm}\sim t^{-2/3}.
\end{eqnarray}
From this estimate, we see that the p--h mode
properly accounts the divergence of the scalar susceptibility
at the Z$_2$CP, and therefore the
softening
of the p--h mode is the origin of the critical divergence
at the Z$_2$CP.

\section{discussions}
\label{sec:4}

In the microscopic calculation with
the NJL model, we have seen that
the collective p--h mode, besides the sigma meson mode, 
is  generated in the scalar channel
and brings  the spectral contributions to
the channels of the conserved quantities through the mixing
when $M\ne 0$.
This p--h spectrum makes the susceptibilities of
$\chi_{\mu\mu}$ and $\chi_{TT}$
discontinuous across the O(4) critical line,
and eventually gives rise to the critical divergences at the
TCP and the Z$_2$CP. This role  of 
the p--h mode is consistent with the behavior of the
hydrodynamic mode in  the TDGL analysis.
We remark here that this p--h mode in the NJL model
is the time--reversible landau--damping type.
In the phenomenological TDGL approach, on the other hand,
we assumed the time--irreversible diffusion motion for
the conserved density, which 
seems more appropriate to the non--equilibrium soft
dynamics.
It would be very interesting to study
how a time--irreversible equation of motion emerges
out of the time--reversible microscopic theory
(see, e.g., Refs.~\cite{BGR98,BC00,HK01,BD03}).

The flat curvature of the effective potential is usually referred to as
the vanishing screening mass, which naively hints the reduction
of a kind of particle mass. 
As we approach the O(4)CP, the sigma meson mass actually gets reduced to 
cause the critical divergence.
However, approaching the Z$_2$CP,
we see that the flat curvature leads to 
the vanishing diffusion constant of the hydrodynamic mode\cite{HF03,FO04,SS04}.
In general, the potential curvature expresses the stiffness
of the system with respect to the variation of the ordering density.
The dynamic quantity  related to this stiffness
can be the particle mass, sound velocity, relaxation constant
or diffusion constant,
depending on the equation of motion of the critical eigenmode.

In the Z$_2$CP case, the linear mixing of the conserved densities in the
proper ordering density dictates that the
critical eigenmode should have hydrodynamic character.
We have seen that the sigma--meson like mode is massive and is
decoupled
from the slow dynamics.
It is explicitly 
argued in Ref.\cite{SS04} that the remaining set of slow modes
is equivalent to that of the liquid--gas critical point\cite{HF03,FO04}.
In the course of heavy ion events passing by the Z$_2$CP,
it is important to study the observable implications of 
critical slowing of hydrodynamic fluctuations in
baryon number and entropy 
densities\cite{RS99,BR00,BPSS01,BG01,PSD03,AFK03,HS03}.
The growth of the diffusive fluctuations
within the finite space--time would get renewed 
interest\cite{RS99,BR00}.

The critical soft mode of the TCP 
is different between the symmetric and broken phases.
The sigma mode becomes the critical eigenmode
and the hydrodynamic one behaves normally
if the critical point is approached from the
exactly symmetric phase.
Otherwise the critical eigenmode bears the hydrodynamic character and
the sigma mode slows down only moderately, which implies the
importance of the hydrodynamic fluctuations near the TCP:
the endpoint of the first order line sitting on the chiral phase boundary.
Theoretically and also experimentally 
it is worthwhile to elaborate and classify the dynamic critical behavior
at the TCP as well as other critical points of 
QCD\cite{FO04,SS04,KO04,KM03,BDW03}.

In the QCD thermodynamics the first order transition is believed
to occur at finite temperature in the massless three--flavor case\cite{PW84}
and in the pure gluonic case\cite{SY82}, where the chiral symmetry and the
$Z_3$ center symmetry are exact, respectively.
As varying masses of the quarks from these two limits,
we will have critical points (line) at the edge of 
the first--order region. 
For example, 
in Ref.~\cite{KLS01}, 
a Z$_2$CP is studied in the $T$--$m$ plane
with three quark flavors of equal mass
in lattice QCD.
There a particular linear combination of the quark
condensate and the energy density
is identified as a proper order parameter, which should be mapped 
to the magnetization in  the Z$_2$ Ising model.
We expect that 
a hydrodynamic mode related to the energy fluctuation
shows critical slowing at this Z$_2$CP, whose spectrum
may be detected in the lattice QCD.
The dynamic critical behavior of this point is also of importance.
It could be different from that of the Z$_2$CP at finite $\mu_B$
because of no linear mixing with the baryon number fluctuation 
due to the symmetry $\rho_B \leftrightarrow -\rho_B$.
Furthermore, the lines and surfaces of the QCD critical points
in the $T$--$\mu_B$--$m_{ud}$--$m_s$ space
are speculated\cite{SA02}.
One can also extend the space to the isospin channel\cite{SS00}.
Generally at such a critical point the proper ordering density
becomes a linear combination involving conserved densities.
Since the critical eigenmode must have the hydrodynamic
character in this case,
the dynamic critical nature 
would be quite different from the case with
(e.g.) the exact chiral symmetry.

Our identification of the soft modes
along the critical line
is done within the mean--field approximation.
Fluctuations around the mean fields
are known to become crucial for describing the
singular behavior at the critical points correctly.
To this end we should utilize 
mode-coupling theory or dynamic renormalization group method.
The mean--field analysis provides
a good starting point to identify an appropriate set of the slow
modes.

\section{Summary}
\label{sec:5}

The fundamental points about the Z$_2$CP are following:
(1) in the absence of the chiral symmetry,
the ordering density becomes  a {\em linear} combination
of the scalar density, the baryon number density and the energy density,
in general, and their susceptibilities have the same critical exponent.
In describing the static property of this critical
point one may equally well take any of these densities as
the ordering density.
(2)
Then the critical eigenmode must be the hydrodynamic one
which can cause the critical divergence of the susceptibility of the
conserved density.  
On the other hand, 
in the chiral transition approached from the symmetric phase,
the exact chiral symmetry
prohibits the linear mixing of the hydrodynamic mode in the fluctuation
of the ordering density.
We have showed these points using the TDGL approach as well
as the microscopic NJL model.

We have studied the change of the critical eigenmode 
along the O(4) critical line. 
When the critical point is approached
from the symmetric phase, the soft mode is indeed
the sigma meson mode. On the other hand, approaching from the
broken phase we see the scalar condensate allows the linear mixing 
between the sigma and hydrodynamic modes, and eventually at the TCP
the hydrodynamic mode turn out to be the critical eigenmode
which generates the leading critical divergence.
Thus the shift of the critical mode from the sigma meson to
the hydrodynamic mode occurs at the TCP. And the 
soft mode at the TCP crucially depends on from 
which phase one is approaching the point.

The criticality of the Z$_2$CP is given by the softening
of a hydrodynamic mode. 
The sigma mode remains as a fast mode 
due to the explicit breaking $m$
and is decoupled from the soft dynamics.
Based on this understanding, we should study
fluctuations with the hydrodynamic character,
such as baryon number and entropy 
fluctuations\cite{RS99,BR00,BPSS01,BG01,PSD03,AFK03,HS03,JK03}
in locating the Z$_2$CP experimentally.

In the QCD thermodynamics with three quark favors,
several kinds of end points are speculated.
One should keep in mind that at these points
the hydrodynamic mode will become the critical eigenmode
once a conserved density is linearly mixed in the
ordering density
and also that the {\em approximate} symmetry
of the underlying interactions does not provide
any reason for singularity in the phase diagram.

\acknowledgments
The authors thank K.~Ohnishi for discussions in
the early stage of this work.
They are very grateful to T.~Matsui, M.~Rho and M.A.~Stephanov 
for fruitful discussions and comments.
An important comment by T.~Kunihiro on
the susceptibility of the conserved density
is sincerely acknowledged.
H.F.\ thanks the warm hospitalities extended to him
at YITP and at KIAS, where parts of this work were performed.
This work is supported in part by the Grants-in-Aid for Scientific
Research of Monka-sho (Grant No.~13440067). 

\appendix

\section{susceptibility and response function}

The susceptibility of a (bosonic) density $\phi_a$ in a system described
with the grand potential 
$ \Omega =-T \ln {\rm tr}(\exp(-\beta \hat K))$
with
$\hat K=\hat H-\sum_a a \int d^3x \hat \phi_a$ is defined as
\begin{eqnarray}
\chi_{ab} \equiv - {1 \over V}
\frac{\partial^2 \Omega}{\partial a \partial b}.
\end{eqnarray}
This susceptibility is obtained as $q$-limit of the response 
function\cite{DF75,TK91}
because
\begin{eqnarray}
\chi_{ab}
&=&\frac{\partial}{\partial a}
e^{\beta \Omega}
tr(\hat \phi_b(0,{\bf 0}) e^{-\beta\hat K})
\nonumber \\
&=&e^{\beta \Omega}
\beta \int _0^1 ds \int d^3 x 
tr(\hat \phi_b(0,{\bf 0}) e^{-s\beta \hat K}
\hat \phi_a(0,{\bf x})e^{-(1-s)\beta \hat K})
-\phi_a \phi_b
\nonumber \\
&=&\int_0^\beta d\tau \int d^3 x \chi^>_{ab}(-i\tau,{\bf x})
=\lim_{\bq \to 0}\chi_{ab}^> (i\omega_n,\bq)|_{n=0}
=\lim_{\bq \to 0}\chi_{ab} (0,\bq),
\end{eqnarray}
where 
we used a formula, 
$\frac{d}{da}e^{A(a)}
=\int_0^1ds e^{sA(a)}\frac{dA(a)}{da}e^{(1-s)A(a)}$
with a matrix--valued function $A(a)$,
and the time dependence of the operators are defined by
$\hat \phi(-i\tau)=e^{\tau \hat K}\hat \phi(0)e^{-\tau\hat K}$.
The imaginary--time correlation and the response function
are introduced as
$\chi_{ab}^>(-i\tau,{\bf x})=
\langle \hat \phi_a(-i\tau, {\bf x}) \hat \phi_b(0,{\bf 0}) \rangle_c$
and 
$\chi_{ab}(t,{\bf x})=
i\theta(t)\langle [ \hat \phi_a(t,{\bf x}), \hat \phi_b(0,{\bf 0})] \rangle_c$,
respectively.
Once we establish the relation between the susceptibility and the
response function, it is easy to express the susceptibility as
an integral over the spectral density:
\begin{eqnarray}
\chi_{ab} = \chi_{ab}(0,\bq \to 0)
=\lim_{\bq \to 0}\int \frac{d \omega}{2\pi}
{2 {\rm Im}\chi_{ab}(\omega, \bq) \over \omega}.
\end{eqnarray}

Specifically, when $\phi_a$ is conserved, $[\hat K, \hat \phi_a]=0$, 
we can freely change the position of 
the operator $\hat \phi_a$ in the trace and
the susceptibility
is directly related to the equal--time correlation function $S$
via\cite{DF75,TK91} 
\begin{eqnarray}
\chi_{ab} &=& \beta \int  d^3x \langle 
\hat \phi_b(0, {\bf 0})
\hat \phi_a(0, {\bf x}) \rangle_c \nonumber \\
&=&
\beta \int  d^3x S_{ab}(0, {\bf x})
=\beta \lim_{\bq \to 0}\int \frac{d\omega}{2\pi}
S_{ab}(\omega, {\bq}) \nonumber \\
&=&\beta \lim_{\bq \to 0}\int \frac{d\omega}{2\pi}
\frac{2 {\rm Im} \chi_{ab}(\omega, {\bq})}{
1-e^{-\beta \omega}}.
\end{eqnarray}
In the last equality we used the fluctuation--dissipation
theorem, which relates the fluctuation $S$ 
to the dissipative part  ${\rm Im}\chi$ of the response function.
Noting the spectral condition,
${\rm sign}(\omega){\rm Im}\chi(\omega,{\bq}) \geq 0$,
we conclude that these two expressions for
the susceptibility coincide if and only if
$\lim_{\bq \to 0}{\rm Im}\chi(\omega,\bq) 
=  \pi  \delta (\omega)\omega\chi$.
Physically this is a consequence of the existence of the
current ${\bf j}_a$ such that 
$\partial _t \phi_a +\nabla \cdot {\bf j}_a=0$.

\section{Explicit expressions}
\subsection{response functions}

We present the explicit formulas and procedures to evaluate
for the one--loop polarization functions.

First working in the imaginary time formalism\cite{LB96}, we derive the
expressions for the polarization functions with make use of the
frequency sum formulas:
\begin{eqnarray}
{\cal I}(i\omega_l) &\equiv&
T\sum_n {1 \over \tilde k_E^2+M^2}{1 \over (\tilde k-q)_E^2+M^2}
\nonumber \\
&=&
{-1 \over 4 E_1 E_2}\left(
{1-n_{+1}-n_{-2} \over  i\omega_l -E_1-E_2}
-{n_{-1}-n_{-2} \over  i\omega_l +E_1-E_2}
+{n_{+1}-n_{+2} \over  i\omega_l -E_1+E_2}
-{1-n_{-1}-n_{+2} \over i\omega_l +E_1+E_2}
\right ).
\\
{\cal I}_\omega(i \omega_l)
&\equiv&
T \sum_n i (2\tilde k_4-q_4)
{1 \over \tilde k_E^2+M^2}{1 \over (\tilde k-q)_E^2+M^2}
\nonumber \\
&=&
{1 \over  4 E_2}\left[
{ 1-n_{+1}-n_{-2} \over i\omega_l- E_1 -E_2}
+
{ n_{-1}-n_{-2} \over i\omega_l+ E_1 -E_2}
+
{ n_{+1}-n_{+2} \over i\omega_l- E_1 +E_2}
+
{ 1-n_{-1}-n_{+2} \over i\omega_l + E_1 +E_2}
\right ]
\nonumber \\
&&
-{1 \over  4 E_1}\left[
{ 1-n_{+1}-n_{-2} \over i\omega_l- E_1 -E_2}
-
{ n_{-1}-n_{-2} \over i\omega_l+ E_1 -E_2}
-
{ n_{+1}-n_{+2} \over i\omega_l- E_1 +E_2}
+
{ 1-n_{-1}-n_{+2} \over i\omega_l + E_1 +E_2}
\right ]
\end{eqnarray}
where $\omega_l=2l \pi T$ is the bosonic Matsubara
frequency, $\tilde k=(\bk,-\omega_n + i\mu)$
the quark momentum with $\omega_n=(2n+1)\pi T$, 
$\tilde k_E^2=\bk^2+\tilde k_4^2$,
$E_1=\sqrt{M^2+\bk^2}$,
$E_2=\sqrt{M^2+(\bk-\bq)^2}$,
and $n_{\pm 1,2}=n_\pm(E_{1,2})$.
The $\bq$-- and $\bk$--dependences are implicit through these
quantities.
Then the analytic continuation 
to the real frequency 
is done by the replacement $ i \omega_l \to q_0+i \varepsilon$,
which uniquely gives the retarded functions
with the asymptotic behavior $\propto 1/q_0$
as $|q_0| \to \infty$.

\begin{figure}[tb]
\centerline{
\hfill
\epsfxsize=0.35\textwidth
\epsffile{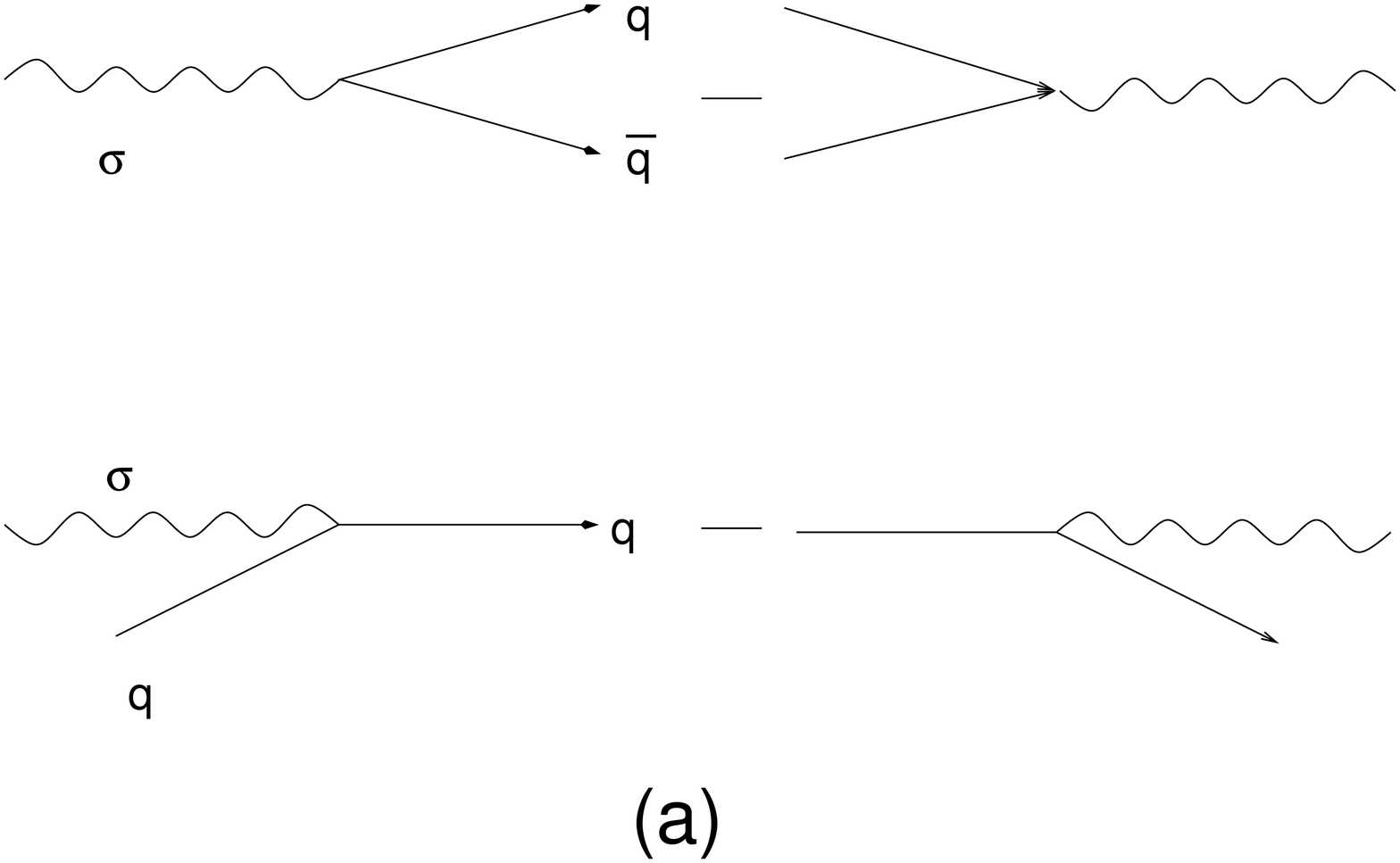}
\hfill
\epsfxsize=0.35\textwidth
\epsffile{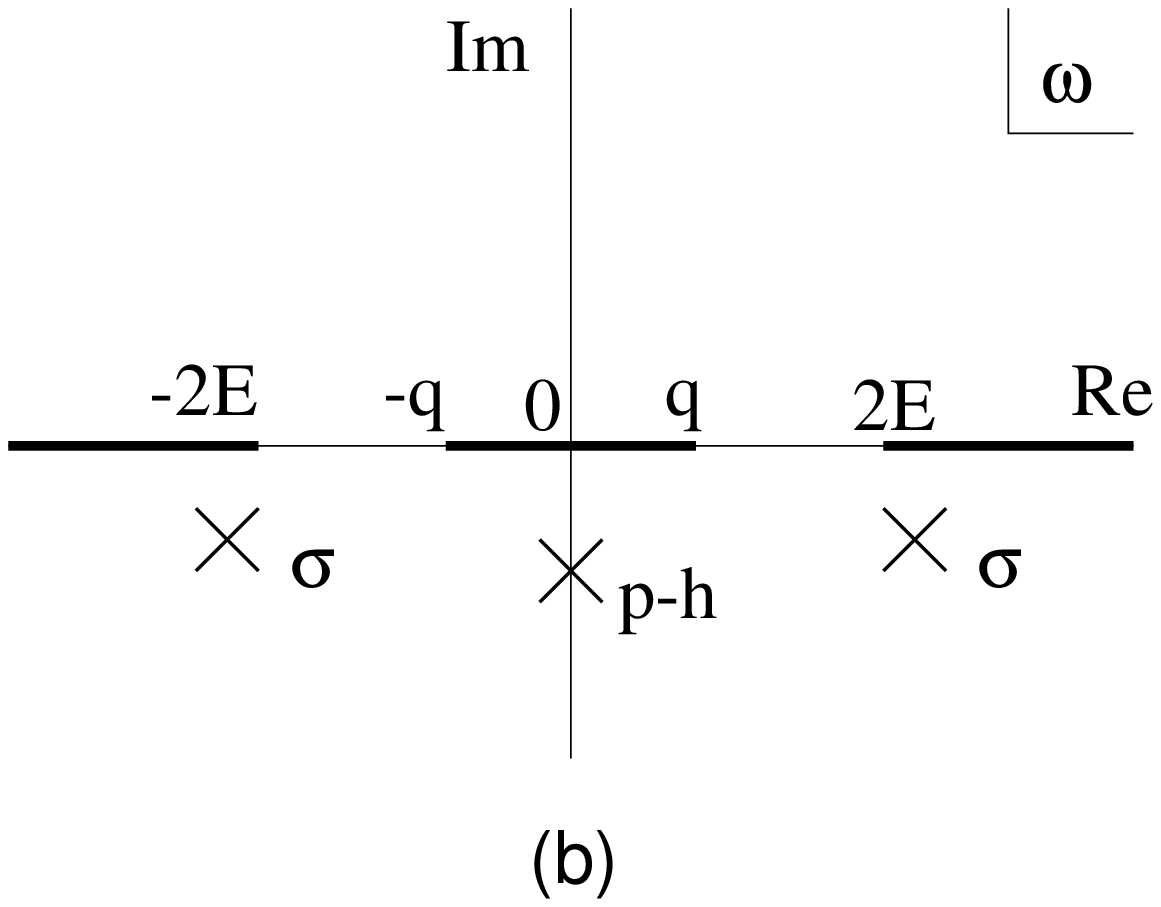}
\hfill
}
\caption{(a)
Pair creation/annihilation (upper) and absorption/emission (lower)
processes contributing to the mode spectrum with the detailed 
balance. (b) Schematic analytic structure of the response function at finite
momentum $q$.
The cut (bold line) on the real axis with $|\omega|>2E$ is due to
the upper processes of (a) while the cut between $\omega = -q$ and $q$
is from the lower processes of (a).
The $\sigma$ meson pole and the p--h pole on the unphysical Riemann sheet
are shown with $\times$.
}
\label{fig:cut}
\end{figure}

Since the retarded function is analytic in the upper half plane,
we can reconstruct it from the imaginary part using the dispersion
integral,
\begin{eqnarray}
\Pi_{ab}(\omega,\bq)={1 \over \pi}\int_{-\infty}^\infty  d\omega'
{{\rm Im}\Pi_{ab}(\omega',\bq) \over \omega' - \omega -i \varepsilon }.
\end{eqnarray}
This relation is quite useful when we evaluate the polarization functions
with finite $\bq$ because the imaginary part is easier to calculate.
The imaginary part comes from two physical processes in our model:
the $q$--$\bar q$ creation/annihilation and the mode-absorption/emission
by a quark $q$ or an anti-quark $\bar q$ as shown in Fig.~\ref{fig:cut}.
Kinematically the former occurs for the time--like momentum 
with $q^2=q_0^2-\bq^2>4M^2$. 
The latter is possible for $q^2 <0$, resulting in the Landau damping.
Then the complex $\omega$ plane has two kinds of cuts in the case
of finite $\bq$.
Analytic continuation of the retarded function
to the lower half plane (${\rm Im} z <0$) across one of these cuts
is defined by
\begin{eqnarray}
\Pi(z,\bq)&=&
{1 \over \pi}\int_{-\infty}^{\infty} d\omega 
{ {\rm Im} \Pi(\omega,\bq) \over \omega -z}
+2{\rm Im} \Pi^{\rm tm,sp}(z,\bq), 
\end{eqnarray}
where $\Pi^{\rm tm,sp}(z,\bq)$ is the imaginary part in the
time-like or the space-like region, depending on across
which cut the function is continued.
Substituting these expressions to Eq.~(\ref{eq:response}), we
can obtain the response functions with finite $\bq$ and search
the poles on the unphysical Riemann sheet (see Fig.~\ref{fig:cut}).

The one--loop polarization function in the scalar channel
yields
\begin{eqnarray}
\Pi_{mm}(q_0,{\bf q}) 
&=&
\nu\int {d^3 k \over (2 \pi)^3}
{1-n_{+1}-n_{-1} \over E_1 }
+
\nu \int {d^3 k \over (2 \pi)^3}
(q^2-4M^2){\cal I}(q_0+i\varepsilon)
\nonumber \\
&\equiv&
J(|\bq|)+(q^2-4M^2)I(q_0+i\varepsilon,\bq),
\label{eq:I}
\end{eqnarray}
where we have shifted the momentum as $\bk \to \bk+\bq/2$ and 
$E_{1,2}=\sqrt{M^2+(\bk\pm\bq/2)^2}$, and then introduce
the cutoff at $k^2_{\rm max}=\Lambda^2-\bq^2/4$ for this new $\bk$.
In the second line we defined the functions $J$ and $I$, whose
massless limits appear in the expansion series
of the NJL effective potential in terms of $\sigma$, Eq.(\ref{eq:NJL0effpot}).
The integral $I^0$ is seemingly divergent logarithmically
in the infrared region, but is actually finite because
$1-n_+^0-n_-^0 \to 0$ as $\bk\to \bzero$.

After performing the angular integration using
the delta function imposed by the on--shell condition,  
the imaginary part of the scalar polarization for $q_0>0$ yields
\begin{eqnarray}
{\rm Im}\Pi_{mm}(q_0,{\bf q})
=
{\nu \over 16 \pi} \int {dk k\over |\bq|}
{q^2 - 4M^2 \over \sqrt{E^2-q^2/4}}
{\cal D}(q_0,q^2),
\end{eqnarray}
\begin{eqnarray}
{\cal D}(q_0,q^2)\equiv
\left \{
\begin{array}{ll}
2-n_{+1}-n_{-2}-n_{+2}-n_{-1}, & \quad 4M^2<q^2<4(M^2+k^2_{\rm max}) \\
n_{+1}-n_{+2}+n_{-1}-n_{-2},   & \quad q^2<0\\
\end{array}
\right .
\end{eqnarray}
where $q^2=q_0^2-\bq^2$ and $E^2=M^2+\bk^2$.
Due to the on-shell condition of the imaginary part,
the quark energies in $n_{\pm}$ 
are set to be
$E_{1,2}={q_0 \over 2}\pm \sqrt{E^2-q^2/4}$ for $q^2>4M^2$
and
$E_{1,2}=\sqrt{E^2-q^2/4}\pm {q_0 \over 2}$ for $q^2<0$.
The last $k$--integration can be done analytically. 
The imaginary
part for $q_0<0$ is obtained by 
${\rm Im}\Pi(-q_0,\bq)=-{\rm Im}\Pi(q_0,\bq)$.

Similarly, other polarization functions and their imaginary parts
for $q_0>0$ are found and given below:

\begin{eqnarray}
\Pi_{\mu\mu}(q_0,{\bf q})&=&
\nu\int {d^3 k \over (2 \pi)^3}
{1-n_{+1}-n_{-1} \over E_1}
+
\nu \int {d^3 k \over (2 \pi)^3}
(q_0^2-4E^2) {\cal I}(q_0),
\\
{\rm Im}\Pi_{\mu\mu}(q_0,{\bf q})
&=&
{\nu \over 16 \pi} \int {dk k\over |\bq|}
{q_0^2-4E^2  \over \sqrt{E^2-q^2/4}}
{\cal D}(q_0,q^2),
\end{eqnarray}
\begin{eqnarray}
\Pi_{m\mu}(q_0,{\bf q})
&=&
-2M \nu \int {d^3 k \over (2 \pi)^3}
{\cal I}_\omega(q_0) ,
\\
{\rm Im}\Pi_{m\mu}(q_0,{\bf q})
 &=&
-{4M\nu \over 16 \pi} \int {dk k\over |\bq|}
{\cal D}_\omega(q_0,q^2),
\end{eqnarray}

\begin{eqnarray}
\Pi_{\beta\beta}(q_0,{\bf q})&=&
\nu\int {d^3 k \over (2 \pi)^3}
{1-n_{+1}-n_{-1} \over E_1}(E^2+\mu^2)
\nonumber \\
&& +
\nu \int {d^3 k \over (2 \pi)^3}
\left [
\left (
(q_0^2-4E^2)(E^2+\mu^2)-E^2 \bq^2+(\bk \cdot \bq)^2
\right ) {\cal I}(q_0)
+\mu (q_0^2 -4 E^2) {\cal I}_\omega (q_0)
\right ] ,
\nonumber \\
\\
{\rm Im}\Pi_{\beta\beta}(q_0,{\bf q})
 &=&
{\nu \over 16 \pi} \int {dk k\over |\bq|}
\left [
{(q_0^2-4E^2)(E^2-q^2/4+\mu^2)
 \over \sqrt{E^2-q^2/4}}
{\cal D}(q_0,q^2)
+2\mu (q_0^2 - 4E^2)
{\cal D}_\omega(q_0,q^2)
\right ]  ,
\end{eqnarray}

\begin{eqnarray}
\Pi_{m\beta}(q_0,{\bf q})
&=&
M \nu\int {d^3 k \over (2 \pi)^3}
{1-n_{+1}-n_{-1} \over E_1}
 +
M \nu \int {d^3 k \over (2 \pi)^3}
\left [
 (q^2-4E^2){\cal I}(q_0)
-2 \mu {\cal I}_\omega (q_0)
\right ]  ,
\\
{\rm Im}\Pi_{m\beta}(q_0,{\bf q)}
 &=&
{M\nu \over 16 \pi} \int {dk k\over |\bq|}
[4\sqrt{E^2-q^2/4}\, {\cal D}(q_0,q^2)
-4\mu {\cal D}_\omega(q_0,q^2)]  ,
\end{eqnarray}

\begin{eqnarray}
\Pi_{\mu\beta}(q_0,{\bf q})
&=&
\nu\int {d^3 k \over (2 \pi)^3}
{1-n_{+1}-n_{-1} \over E_1} \mu
\nonumber \\
&& +
\nu\int {d^3 k \over (2 \pi)^3}
[\mu(q_0^2-4E^2+\bq^2){\cal I}(q_0)
+{1 \over 2}(q_0^2 -4E^2){\cal I}_\omega(q_0)]  ,
\\
{\rm Im}\Pi_{\mu\beta}(q_0,{\bf q})
 &=&
{\nu \over 16 \pi} \int {dk k\over |\bq|}
\left [
\mu { q_0^2-4 E^2+\bq^2 \over \sqrt{E^2-q^2/4}}\, {\cal D}(q_0,q^2)
+(q_0^2-4E^2) {\cal D}_\omega(q_0,q^2)
\right ].
\end{eqnarray}

In these expressions we introduced a function
\begin{eqnarray}
{\cal D}_\omega(q_0,q^2)
\equiv
\pm 
(n_{+1}+n_{-2}-n_{-1}-n_{+2})
\end{eqnarray}
with the sign `$+$' for $4M^2<q^2<4(M^2+k^2_{\rm max})$
and `$-$' for $q^2<0$.

\begin{figure}[tb]
\centerline{
\hfill
\epsfxsize=0.25\textwidth
\epsffile{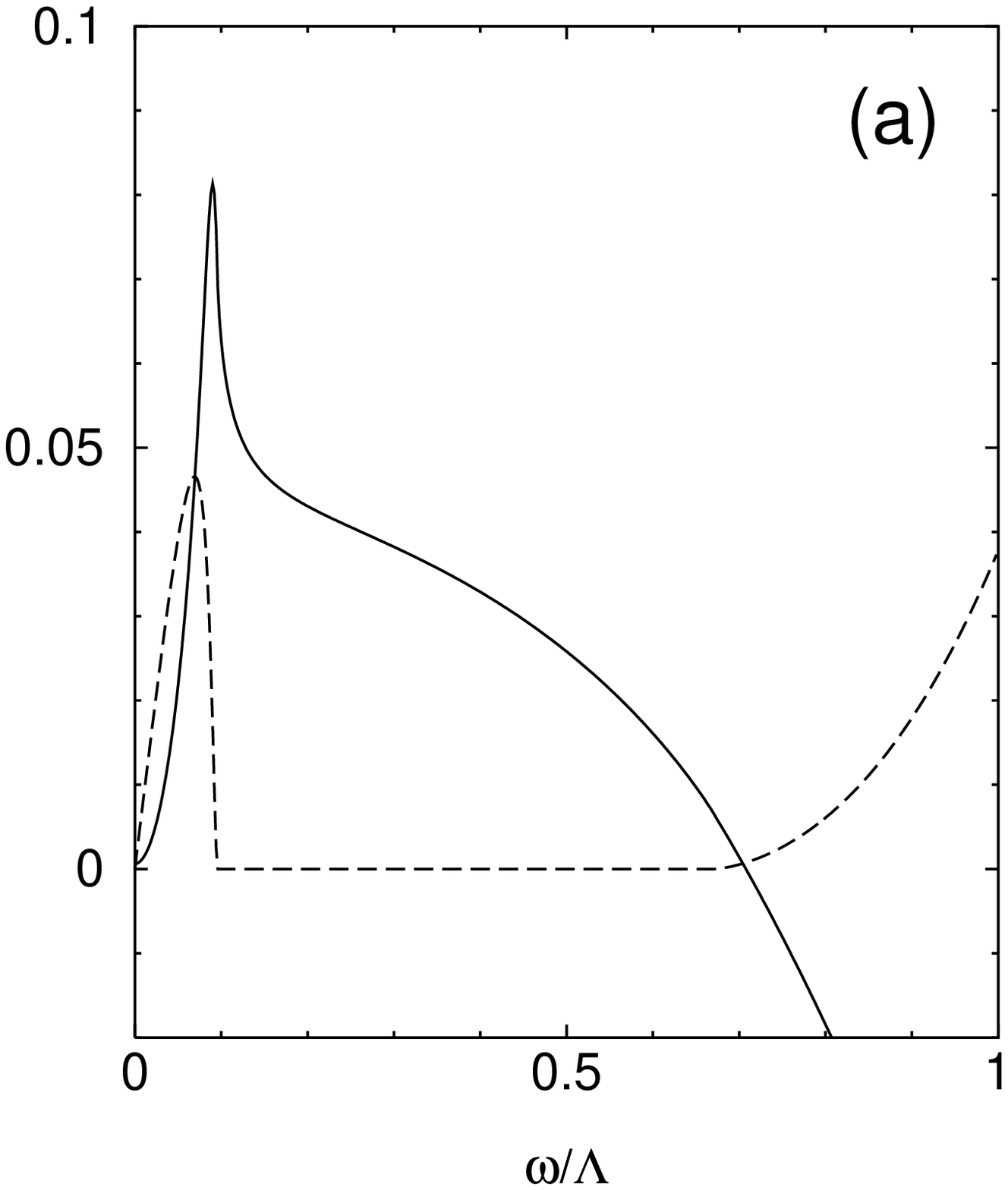}
\hfill
\epsfxsize=0.37\textwidth
\epsffile{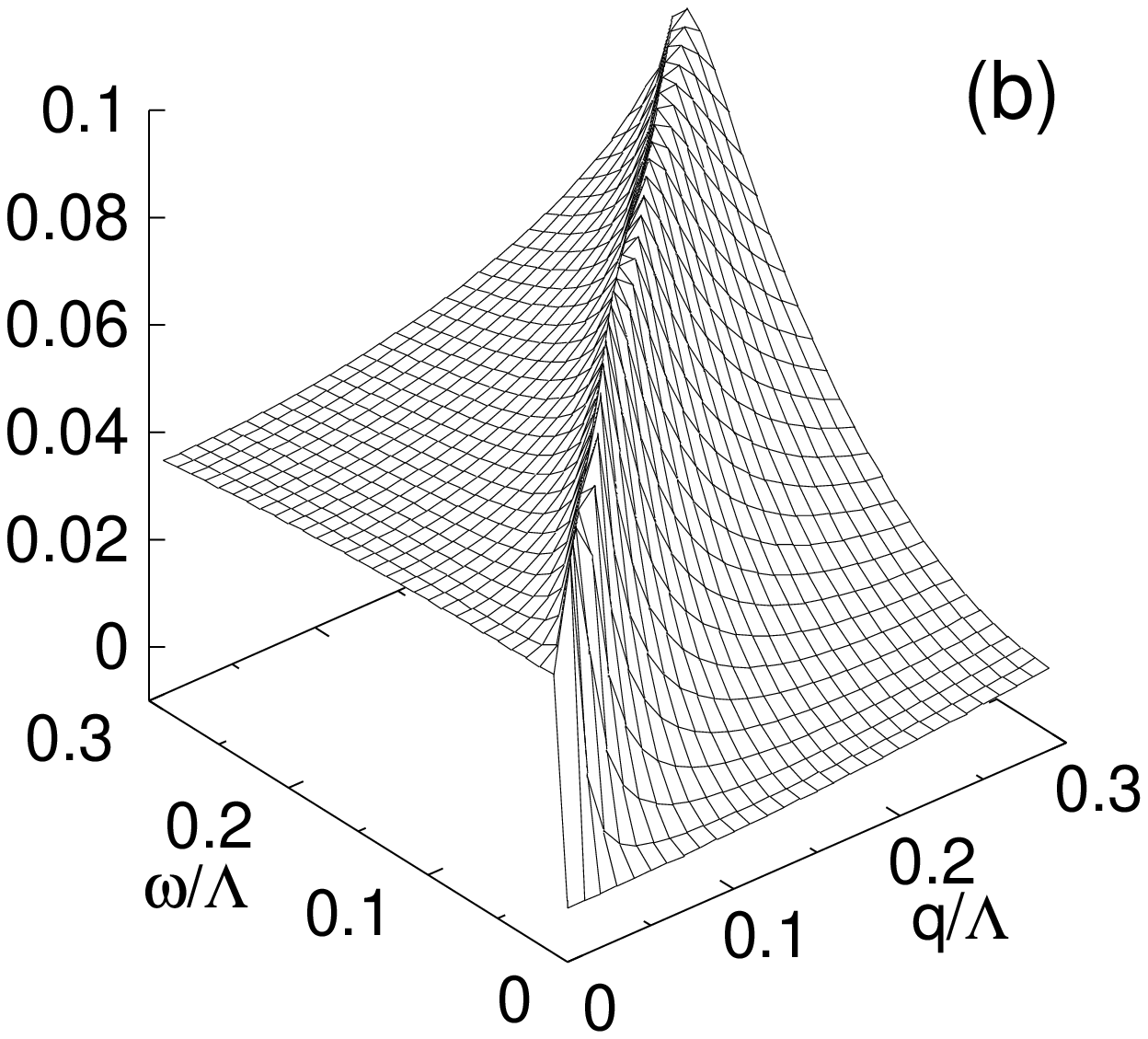}
\hfill
\epsfxsize=0.37\textwidth
\epsffile{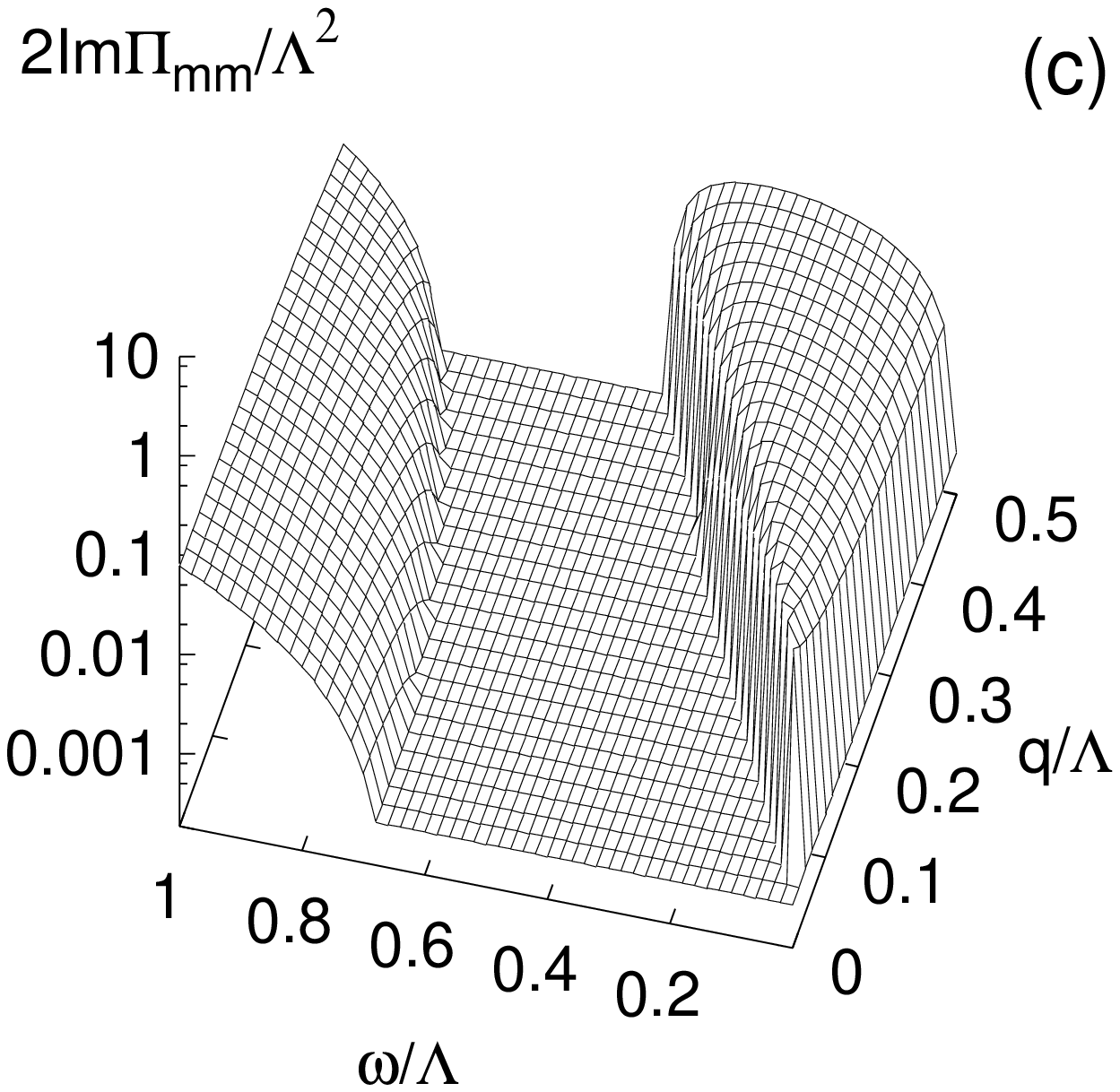}
\hfill
}
\caption{One--loop scalar polarization function $\Pi_{mm}$
(at the Z$_2$CP).
(a) The real part $1/(2g)-{\rm Re}\Pi_{mm}$  and
the imaginary part Im$\Pi_{mm}$ are shown
in solid and dashed lines, respectively, with fixed $q/\Lambda=0.1$.
(b) The real part as a function of $(\omega, q)$.
(c) The imaginary part as a function of $(\omega, q)$.}
\label{fig:singular}
\end{figure}

\begin{figure}[tb]
\centerline{
\hfill
\epsfxsize=0.37\textwidth
\epsffile{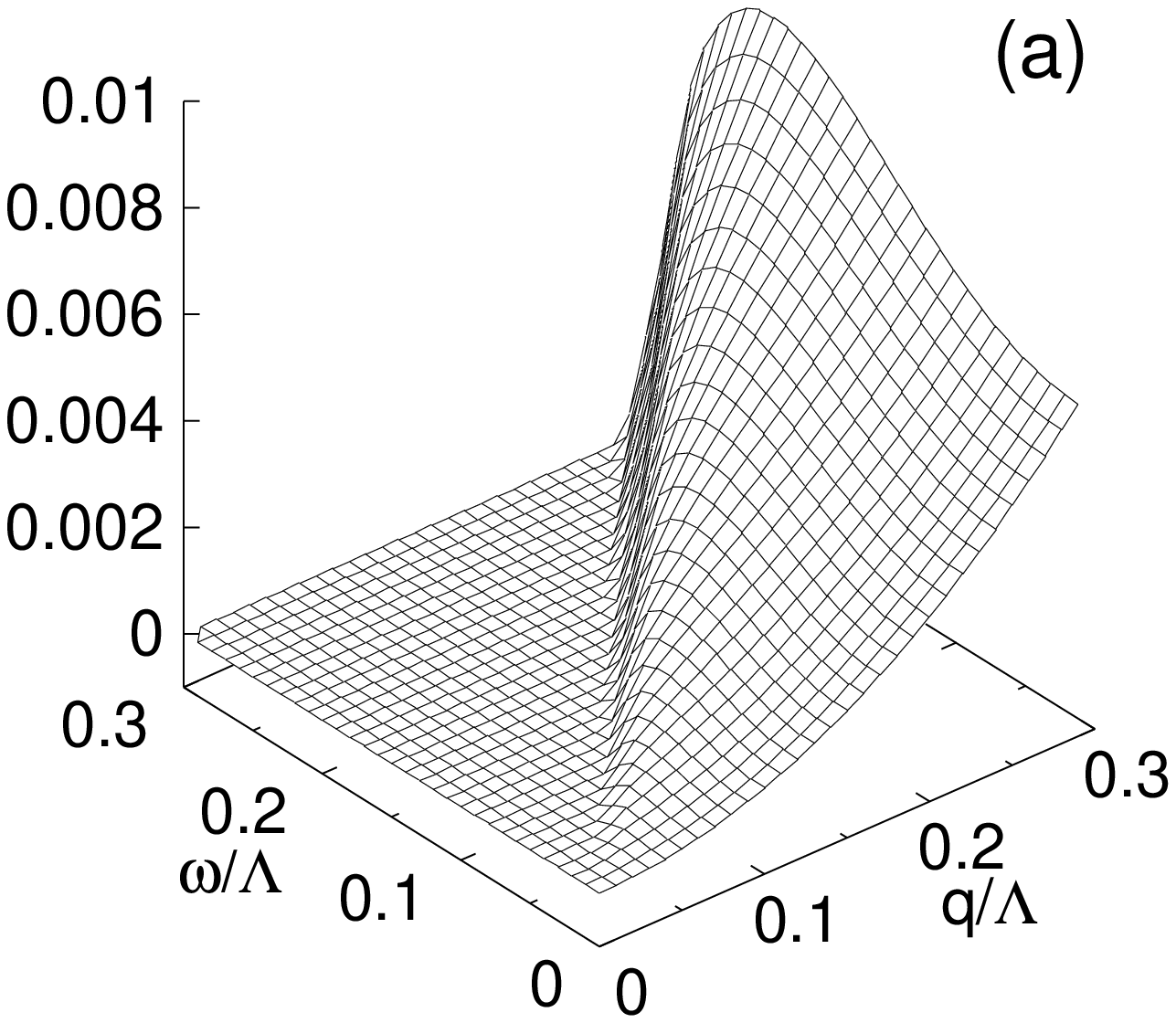}
\hfill
\epsfxsize=0.37\textwidth
\epsffile{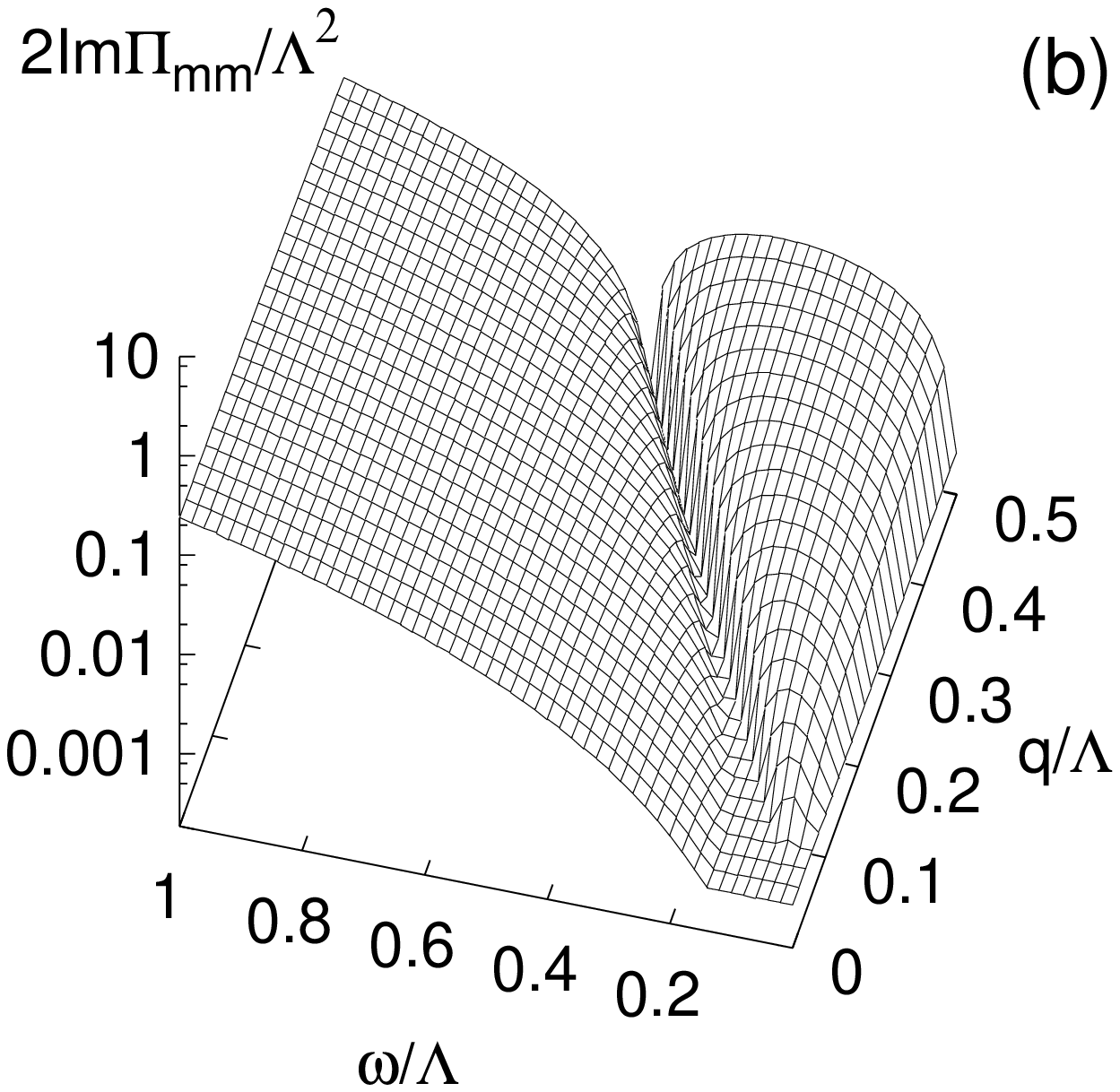}
\hfill
}
\caption{One--loop scalar polarization function
with massless quarks (at the TCP):
(a) the real part $1/(2g)-{\rm Re}\Pi_{mm}$ and (b) the imaginary part
Im $\Pi_{mm}$.}
\label{fig:scalarpol}
\end{figure}

It is known that the
response functions have a non--analytic property at the origin of the
$\omega$--$q$ plane when the p--h mode spectrum exists. 
For demonstration we show in Fig.~\ref{fig:singular} 
the real and imaginary parts of
$\Pi_{mm}(\omega,\bq)$ with $\bq/\Lambda =0.1$.
An abrupt change of the real part is seen 
in the region $\omega/\Lambda <0.1$, which 
is clearly caused by the p--h spectrum in accord with
the dispersion relation. 
In the $\bq \to \bzero$ limit, the imaginary
part becomes proportional to $\omega \delta (\omega)$, which
leads to the discontinuity of the real part of $\Pi_{mm}$
as mentioned in Eq.~(19) and shown in Fig.~\ref{fig:singular}. 
In the case of the massless quarks, on the other hand,
the scalar channel does not couple to the p--h motion
in the $\bq \to \bzero$ limit due to the chiral symmetry.
Hence the real part is non-singular at the origin as shown in
Fig.~\ref{fig:scalarpol}.

In Appendix A it is shown that 
the spectral function for a conserved density fluctuation
must be proportional to $\omega \delta(\omega)$
as $\bq \to \bzero$ in general.
We confirm that this requirement
is fulfilled by these one-loop polarizations,
$\Pi_{\mu\mu}$, 
$\Pi_{m\mu}$,
$\Pi_{\beta\beta}$, and
$\Pi_{m\beta}$.

\subsection{susceptibilities}

The susceptibilities in free quark gas with mass $M$ are found as
\begin{eqnarray}
\chi_{mm}^{(0)}&=&
\nu \int {d^3 k\over (2\pi)^3}\left [ {1\over E}(1-n_+ -n_-)
-{M^2 \over E^2}
\left ({1-n_+ - n_- \over E}+
n'_+ + n'_- \right ) \right ]  
,
\nonumber \\
\chi_{m\mu}^{(0)}&=&\nu \int\frac{d^3k}{(2\pi)^3}\frac{M}{E}(n_+'-n_-')
,
\nonumber \\
\chi_{mT}^{(0)}&=&\nu \int\frac{d^3k}{(2\pi)^3}\frac{M}{E}
\left [\frac{E-\mu}{T} n_+' +\frac{E+\mu}{T} n_-' \right ]
,
\nonumber \\
\chi_{\mu\mu}^{(0)}&=&-\nu \int\frac{d^3k}{(2\pi)^3}(n_+'+n_-')
,
\nonumber \\
\chi_{TT}^{(0)}&=&-\nu \int\frac{d^3k}{(2\pi)^3}\left [
 \left (\frac{E-\mu}{T}\right )^2 n_+'
+\left (\frac{E+\mu}{T}\right )^2 n_-' \right ]
,
\nonumber \\
\chi_{T\mu}^{(0)}&=&-\nu \int\frac{d^3k}{(2\pi)^3}
\left [\frac{E-\mu}{T} n_+' -\frac{E+\mu}{T} n_-' \right ]
.
\nonumber \\
\end{eqnarray}
These expressions coincide with the static one-loop 
polarizations in the $\bq \to \bzero$ limit.
Through this limiting procedure we find that
the terms containing $n_{\pm}'$ are related with
the p--h spectrum parts in
functions ${\cal I}$ and ${\cal I}_\omega$,
and that all susceptibilities, except for the
scalar, must accompany $n_{\pm}'$ because of their
hydrodynamic nature.

We also notice the fact of no mixing of
the scalar fluctuation with others 
in the massless quark gas $M=0$.
In the case of $\mu=0$, the vector fluctuation does not
mix with others.
Both of these originate from the symmetries.
If we write down the
GL effective potential, it must be invariant
under $\sigma \to -\sigma$ and/or $\rho \to -\rho$, respectively,
and therefore any linear mixing with other
densities is impossible.

\section{Chiral quark model}

The chiral quark model can be used to perform
the same analysis as in the NJL model:
\begin{eqnarray}
{\cal L}_{\chi q} &=& \frac{1}{2} (\partial _\nu \phi_\alpha)^2 
-\frac{1}{2} \hat m ^2 \phi_\alpha ^2
-\frac{\lambda}{4!} (\phi_\alpha^2)^2 +h \sigma
+\bar q [i \!\!\not\! \partial - g(\sigma +i\gamma_5 \tau_a \pi_a)]q 
,
\end{eqnarray}
where $\phi_0=\sigma$, $\phi_a=\pi_a$, and $\hat{m}^2 <0$. 
The meson mode is introduced here as an elementary field
with the kinetic term.
Integrating out the quark field,
we obtain the effective potential
within the mean field approximation for $\sigma$ and $\pi$ as
\begin{eqnarray}
\Omega_{\chi q}(T,\mu;\sigma)/V&=&
-h \sigma  
+\frac{1}{2} \hat m ^2 \sigma^2
+\frac{\lambda}{4!} \sigma^4
-\nu \int \frac{d^3 k}{(2\pi)^3}
[E-T\ln (1-n_+)-T\ln(1-n_-)]
\label{eq:chiqfree}
\end{eqnarray}
with $E=\sqrt{M^2+\bk^2}$ and $M=g\sigma$. 
This potential is almost the same as that of the NJL model 
(\ref{eq:NJLfree}), and this model 
is expected to have the same phase structure.
The subtle point is that 
the divergent vacuum quark fluctuation in the integrand
of Eq.~(\ref{eq:chiqfree}) requires a 
regularization and renormalization.
Instead of the three momentum cutoff used in our NJL model calculation,
here we adopt a simple prescription following
(e.g.)~Ref.~\cite{SMMR01};
we assume that the renormalization is already done in the vacuum
and discard the vacuum polarization term.
Then the parameters are chosen
so as to reproduce the pion decay constant, the pion and sigma masses,
and the constituent quark mass in the vacuum.
We found the Z$_2$CP at $(T_c,\mu_c)$=(117.7, 176.2) MeV.

Within the same level of the approximation,
the scalar response function is calculated as
\begin{eqnarray}
\chi_{h}={1 \over -q^2 +\hat m^2+\frac{1}{2}\lambda \sigma^2-g^2 \Pi_{mm}} ,
\end{eqnarray}
where the polarization $\Pi_{mm}$ is defined in Eq.~(\ref{eq:polfuncs})
with $\Gamma=1$,
but whose vacuum part is removed.
The four--point interaction of the NJL model is replaced
by the non--local one here.
Other response functions
\begin{eqnarray}
\chi_{ab}=\Pi_{ab}+\Pi_{am}
{g^2 \over -q^2 +\hat m^2+\frac{1}{2}\lambda \sigma^2-g^2 \Pi_{mm}}
\Pi_{mb} \qquad (a,b=\mu,\beta)
\end{eqnarray}
have the same structure as the NJL result (\ref{eq:response})
because we assume the same scalar--type  interaction between quarks.

In the numerical calculation with this chiral quark model at the Z$_2$CP,
we confirmed the spectral enhancement in the space--like
momentum region, similar to Fig.~\ref{fig:spfunc},
and found a pole responsible for this enhancement on the
negative imaginary axis in the complex--$\omega$ plane, just as
in the NJL model.
The ratio $R$ defined in Eq.~(\ref{eq:R-NJL})
also goes to unity as the Z$_2$CP approached.
Therefore our conclusion on the importance of the hydrodynamic mode
at the Z$_2$CP is unaltered here.

We should note, however, 
that the semi--positivity condition on 
the spectral function is violated in
our numerical result 
in the time--like momentum region.
This is because 
we replaced the term  $1-n_{\pm 1}-n_{\mp 2}$ 
in the expression of $\Pi_{mm}$ with $-n_{\pm 1}-n_{\mp 2}$
to remove the divergence (see Appendix B). 
This simple regularization
breaks the detailed balance relation which is
essential to assure thermal equilibrium.
Hence the result of the spectrum in this chiral quark model
should be interpreted with caution.
The vacuum subtraction also results in the unexpected
infra--red divergence of the quartic term, $-\lambda +3 g^4 I^0$, 
in the expansion of (\ref{eq:chiqfree}) around $\sigma=0$.
Because of this difficulty,
we could not find the TCP in this model
with the regularization adopted here.
In order to properly discuss the spectral  structure in the chiral quark
model we need the regularization scheme which
satisfies the condition of thermal equilibrium\cite{SMMR01,JPSS04}.

\end{document}